\documentclass[%
 amsmath,amssymb,
 aps,
 pre,
 twocolumn,
 superscriptaddress,
 showpacs
]{revtex4-2}

\usepackage{graphicx}
\usepackage{dcolumn}
\usepackage{bm}
\usepackage{mathtools}
\usepackage{pgfplots}
\usepackage{tikz}
\usepackage{makecell}
\usepackage{multirow}
\usepackage{xcolor}

\usepgfplotslibrary{groupplots}
\pgfplotsset{every tick label/.append style={font=\tiny}}

\newcommand{\pfrac}[2]{\frac{\partial #1}{\partial #2}}
\def\checkmark{\tikz\fill[scale=0.4](0,.35) -- (.25,0) -- (1,.7) -- (.25,.15) -- cycle;}

\definecolor{apsblue}{RGB}{16, 38, 148}
\usepackage[colorlinks=true, allcolors=apsblue]{hyperref}

\newcommand\bp{{\bf p}}
\newcommand\br{ {\bf r}}
\newcommand\bv{{\bf v}}
\newcommand\bfm{{\bf m}}
\newcommand\bQ{{\bf Q}}
\newcommand\tf{{t_{\rm f}}}

\newcommand\bfint{{\mathbf{f}_{\rm int}}}

\newcommand\bu{ {\bf u}}
\newcommand\bfeta{ {\boldsymbol{\eta}}}

\newcommand\bF{{\bf F}}
\newcommand\U{U_{\rm int}}

\newcommand\tU{\tilde{U}_{\rm int}}
\newcommand\eff{{\rm eff}}
\newcommand\ri{ {\rm i} }
\newcommand\bP{{\bf P}}
\newcommand\rf{ {\rm f} }
\newcommand\bJ{{\bf J}}
\newcommand\subA{{ \rm A}}

\newcommand\subIK{{ \rm IK}}
\newcommand\subAOUP{{ \rm AOUP}}
\newcommand\bsigma{{\boldsymbol{\sigma}}}

\begin{document}

\preprint{APS/123-QED}

\title{Exceptions to the Ratchet Principle in active and passive stochastic dynamics}

\author{Jessica Metzger}
\affiliation{Department of Physics, Massachusetts Institute of Technology, Cambridge, Massachusetts 02139, USA}
\author{Sunghan Ro}
\affiliation{Department of Physics, Harvard University, Cambridge, Massachusetts 02138, USA}
\affiliation{Department of Physics, Massachusetts Institute of Technology, Cambridge, Massachusetts 02139, USA}
\author{Julien Tailleur}
\affiliation{Department of Physics, Massachusetts Institute of Technology, Cambridge, Massachusetts 02139, USA}

\date{\today}

\begin{abstract}
The ``ratchet principle" asserts that non-equilibrium systems which violate parity symmetry generically exhibit steady-state {directed} currents. {We study} exceptions to this principle {for stochastic systems in the presence of asymmetric fluctuation sources and show how effective momentum conservation prevents the emergence of {directed} currents. }
For underdamped and overdamped Brownian dynamics, we show {that} thermal fluctuations cannot power the momentum sources required to sustain {directed} currents, even when time-reversal symmetry is broken due to an inhomogeneous temperature field.
While Active Brownian and Run-and-Tumble particles display interaction-induced {directed} currents in asymmetric activity landscapes, we show that {effective momentum conservation prevents this in Active Ornstein-Uhlenbeck} particles:  not all inhomogeneous active fluctuations {can power transport}.
For each of the systems considered in this article, we numerically test for the emergence of interaction-induced {directed} currents. We then characterize time-reversal (a)symmetry in position space using a combination of path-integral and operator methods. When the existence of effective momentum conservation is ruled out, we develop perturbation theories to characterize the onset of interaction-induced {directed} currents.
\end{abstract}

\maketitle

\tableofcontents{}

\section{Introduction}

\begin{figure*}
    \centering
    \begin{tikzpicture}
    \path (-0,0) node {\includegraphics{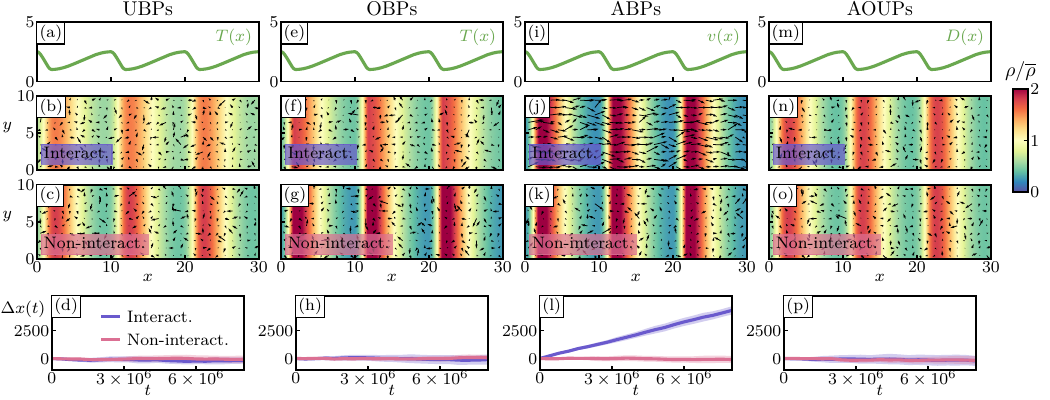}};
    \end{tikzpicture}
    \caption{{Spatially-inhomogeneous fluctuations cause interaction-induced currents in ABPs [{\bf (i-l)}], but not in {UBP}s [{\bf (a-d)}], {OBP}s [{\bf (e-h)}], or AOUPs [{\bf (m-p)}]. In panels (a), (e), (i), and (m), we show the fluctuation landscape used in each system. In panels (b), (f), (j), and (n), we show the steady-state density $\rho$ (heatmap) and current field (arrows) for the systems in the presence of interactions. In panels (c), (g), (k), and (o), we do the same for the non-interacting systems. In panels (d), (h), (l), and (p), we show the average displacement per particle both in the presence (purple) and absence (pink) of interactions. }Simulation details are provided in Appendix~\ref{appendix:sim-examp}.}
    \label{fig:ratchet-examp}
\end{figure*}

The ``Ratchet Principle" states that the violation of time-reversal symmetry (TRS) and parity symmetry generically leads to the emergence of steady-state {directed} currents~\cite{denisov_tunable_2014,jorge}. 
{The term `ratchet'} stems from studies by Smoluchowski, Feynman and others on the rectification of fluctuations into steady motion using ratchet-like {asymmetric potentials~}\cite{smoluchowski1912experimental,feynman_feynman_1963,parrondo1996criticism,reimann_brownian_2002}{. Since then, the scope of the ratchet principle has significantly broadened. Indeed, {directed} currents have been shown to emerge from diverse forms of parity-symmetry breaking, beyond the sole case of asymmetric potentials~}\cite{kidoaki_rectified_2013,caballero_ratchetaxis_2015,rein_force-free_2023,borsley_molecular_2024}{, and we generically refer to such currents as `ratchet currents'.}
Ratchets have found applications in a broad range of topics ranging from the motion of molecular motors~\cite{ajdari_mouvement_1992,rousselet_directional_1994,ajdari_rectified_1994,julicher1997modeling,reimann_brownian_2002,frey2005brownian,campas2006collective,hanggi_artificial_2009} to the rectification of bacterial suspensions~\cite{galajda_wall_2007,angelani_active_2011} and have been proposed as a mechanism to power nonequilibrium microscopic motors~\cite{di2010bacterial,sokolov2010swimming}. {In stochastic systems}, they emerge {from noises and damping that} do not satisfy a fluctuation-dissipation theorem~\cite{magnasco_forced_1993,magnasco_molecular_1994} and are thus {frequently encountered} in active matter~\cite{fiasconaro_active_2008,angelani_active_2011,pototsky_rectification_2013,ghosh2013self,lopatina_self-driven_2013,ai_rectification_2013,reichhardt_active_2013,koumakis_directed_2014,yariv_ratcheting_2014, bijnens_pushing_2021,martin_statistical_2021,obyrne_time_2022,derivaux_rectification_2022,zhen_optimal_2022,khatri_inertial_2023,muhsin_inertial_2023,rojasvega_mixtures_2023,ryabov_mechanochemical_2023,schimming_active_2024,anand_transport_2024,wang_spontaneous_2024}.

Interestingly, exceptions to the ratchet principle have been reported. 
{Some systems, despite being out of thermal equilibrium, indeed admit symmetries that make opposite values of the current equally likely, which rules out a non-zero average current~\cite{denisov_tunable_2014}. 
For ratchets in the presence of external potentials,
this can take the form of transformations of space and time coordinates that relate trajectories with opposite values of the current~}\cite{flach_directed_2000,denisov_tunable_2014,cubero_hidden_2016}{.}

{Stochastic systems with asymmetric, spatially-varying fluctuation sources offer another case of seemingly nonequilibrium systems that lack {directed} currents~}\cite{van_kampen_diffusion_1988,schnitzer_theory_1993,cates_when_2013,martin_statistical_2021}{. A first example is that of overdamped passive Brownian particles ({OBP}s)} in inhomogeneous asymmetric temperature fields~\cite{van_kampen_diffusion_1988,van_kampen_relative_1988}. {There, van Kampen showed that the rectification of inhomogeneous thermal fluctuations into directed motion requires the presence} of an external potential. 
In active matter, run-and-tumble particles (RTPs) and active Brownian particles (ABPs) with spatially varying activity also do not exhibit steady-state currents {in the absence of external potentials~}\cite{schnitzer_theory_1993,tailleur_statistical_2008,cates_when_2013}, a result that has been generalized to active Ornstein-Uhlenbeck particles (AOUPs) {in 1d~}\cite{martin_statistical_2021}. {For asymmetric fluctuation landscapes, no reparametrization symmetry~\cite{denisov_tunable_2014} has been brought forward to explain the lack of {directed} current. Furthermore, introducing} translational diffusion~\cite{rein_force-free_2023}, a \textit{symmetric} external potential~\cite{pototsky_rectification_2013}, {or interactions~\cite{stenhammar_light-induced_2016}} into a system of ABPs with asymmetric activity landscape all lead to the emergence of {directed} currents. {This last result is particularly interesting given the existing interest in interaction-induced ratchets~}\cite{reimann_brownian_2002,van_der_meer_spontaneous_2004,liebchen_interaction-induced_2012,liebchen_interaction_2015}{.}

\begin{table*}[]
    \centering
\renewcommand{\arraystretch}{1.65} 
    \centering
    \begin{tabular}{c|c|c||c|c||c}
        System & \makecell{Spatially varying\\fluctuation source} & \,\makecell{Interactions}\, & \,\makecell{Time-reversal \\symmetry {in}\\ {position space}}\, & \,\makecell{Effective momentum \\conservation}\, & \makecell{{Directed} current}\\
        \hline
        \multirow{2}{*}{\makecell{Underdamped Passive Brownian \\ Particles ({UBP}s)}} & \multirow{2}{*}{\makecell{Temperature}}  &  & & \checkmark & \\
        \cline{3-6}  
        & &  \checkmark & & \checkmark & \\
        \hline
        \multirow{2}{*}{ \makecell{Overdamped Passive Brownian \\
        Particles ({OBP}s)}} & \multirow{2}{*}{\makecell{Temperature}}  &  & \checkmark & \checkmark &  \\
        \cline{3-6}  
        & &  \checkmark & & \checkmark &  \\
        \hline
        \multirow{2}{*}{\makecell{Active Brownian/Run-and-Tumble \\Particles (ABPs/RTPs)}} & \multirow{2}{*}{\makecell{Activity}}  &  & \checkmark & & \\
        \cline{3-6}  
        & &  \checkmark & & & \checkmark\\
        \hline
        \multirow{2}{*}{\makecell{Active Ornstein-Uhlenbeck\\ Particles (AOUPs)}} & \multirow{2}{*}{\makecell{Activity}} &  & & \checkmark &  \\
        \cline{3-6}  
        & &  \checkmark & & \checkmark &
    \end{tabular}
    \caption{Exceptions to the ratchet principle include Underdamped Passive Brownian Particles ({UBP}s) and Overdamped Passive Brownian Particles ({OBP}s) in spatially-varying temperature fields; along with Active Brownian Particles (ABPs), Run-and-Tumble Particles (RTPs), and Active Ornstein-Uhlenbeck Particles (AOUPs) with spatially-varying activity. A {directed} current is prevented in these systems either by a hidden time-reversal symmetry (TRS) or a bulk momentum conservation. In ABPs and RTPs, a {directed} current is induced by adding a symmetric pairwise interaction potential $\U$ between particles, since this destroys any form of TRS or momentum-conservation in the system. (In Section~\ref{sec:od-pbp-whatever}, we show that this is also the case for {OBP}s with time-discretization $\alpha\in (0,1)$.)
    }
    \label{tab:exceptions-summary}
\end{table*}

{Until recently, what controls the existence of ratchet exceptions in asymmetric fluctuation landscapes, and what causes the reemergence of currents, had remained elusive.}
In our companion Letter~\cite{metzger_exceptions_letter}, we filled this gap and showed how {these} ratchet exceptions can be rationalized by amending the ratchet principle. 
{The emergence of directed} currents was shown to require three conditions, not two: the lack of {any form of} time-reversal and parity symmetries, as expected{~\cite{denisov_tunable_2014}}, but also the existence of net momentum sources. {Indeed, we showed that an effective conservation law for the momentum field prevents extraction of the net momentum needed to power a {directed} current. We note that, in some active systems~\cite{o2022time}, irreversibility vanishes upon coarse-graining, leading to large scale physics akin to equilibrium dynamics. This is not the case for the interacting systems considered in our work, and it still would not explain the exact vanishing of the current at the microscopic scale.}
Our results were illustrated by contrasting {OBPs} on the one hand, {where the lack of {directed} current was shown to survive interactions, }and RTPs and ABPs on the other hand{, where interaction-induced currents have been reported~}\cite{stenhammar_light-induced_2016,metzger_exceptions_letter}. 

In this article, we study these systems separately for their own sake, detailing all relevant methods. {Figure~\ref{fig:ratchet-examp} illustrates the setup that we consider across (most) models: an asymmetric fluctuation landscape along $\hat x$, with or without interactions between particles, in which we detect the emergence of a steady {directed} current $\bJ=J \hat x$ or lack thereof. 
In addition to the cases studied in our Letter~\cite{metzger_exceptions_letter}, we} show how the phenomenology of AOUPs and underdamped {passive Brownian particles ({UBP}s)} can also be analyzed within the same framework. 
Our results reveal that the fragility of ratchet exceptions is a subtle question: interactions indeed do not suffice to induce {a {directed} current} in AOUPs with spatially varying activity, at odds with the ABP result. Interactions and activity are thus not sufficient to induce {directed} currents. We consider {UBP}s in Section~\ref{sec:ud-pbps}, {OBP}s in Section~\ref{sec:od-pbps}, ABPs and RTPs in Section~\ref{sec:abp-rtp}, and AOUPs in Section~\ref{sec:aoups}.
In each {of Sections~\ref{sec:ud-pbps}-\ref{sec:aoups}}, we first discuss the existence of TRS {in position space} and then show how, when it is violated, {directed} currents only emerge when there is no effective momentum-conservation in the bulk. {We note that we restrict our study of TRS to position space. As such, the existence of such a form of TRS does not imply a vanishing production rate of thermodynamic entropy or heat dissipation through other degrees of freedom such as momentum or self-propulsion~}\cite{celani_anomalous_2012,shankar2018hidden,o2022time}{. Nevertheless, because we are interested in spatial currents, this definition is sufficient} for our purpose. Our results are summarized in Table~\ref{tab:exceptions-summary}. {Finally, we consider in Sec.~\ref{sec:blowtorch} a ``blowtorch ratchet"~}\cite{buttiker_transport_1987,landauer_motion_1988,blanter_rectification_1998,benjamin_inertial_2008}{, where {OBP}s experience a spatially-varying temperature and an external potential, hence going beyond the framework of Fig.~\ref{fig:ratchet-examp}. There, we show how the inhomogeneous temperature allows rectification of the momentum sources due to the potential, which otherwise balance out for uniform temperature fields. This section thus  explains the emergence of currents in the blowtorch ratchet using a mechanical perspective.} All numerical methods are detailed in appendix~\ref{app:sim}.


\section{Underdamped Passive Brownian Particles}\label{sec:ud-pbps}

We begin by considering {UBP}s, whose positions $\br_i$ and momenta $\bp_i$ evolve according to:
\begin{align}
    \dot{\br}_i &= \bp_i \, ,\label{eq:langevin-ud-pbp-r}\\
    \dot{\bp}_i &= - \gamma \bp_i - \sum_j \nabla \U(\br_i-\br_j) + \sqrt{2 \gamma T(\br_i) } \bfeta_i\;.
    \label{eq:langevin-ud-pbp}
\end{align}
Here the $\{\bfeta_i(t)\}$ are centered  Gaussian white noises that satisfy $\langle \eta_{i,\mu}(t) \eta_{j, \nu}(t') \rangle = \delta_{ij} \delta_{\mu \nu} \delta(t - t')$, with $i$ and $j$ being particle indices, and  $\mu$ and $\nu$ being spatial indices. While our results hold for general potential $\U$, in this article we use
\begin{equation}
    \U(\br)=\frac{\varepsilon \sigma}{2}\bigg(1-\frac{|\br|}{\sigma}\bigg)^2 \Theta(\sigma-|\br|)\;,
    \label{eq:Uint-def}
\end{equation}
where $\Theta$ is the Heaviside function. We denote by $\bfint(\br)=-\nabla \U$ the corresponding interparticle force.

We define the fluctuating phase-space density as
\begin{align}
    \hat{\psi}(\br,\bp,t) &\equiv \sum_i \delta[\br-\br_i(t)] \delta[\bp-\bp_i(t)]
    \label{eq:fluct-P-r-p-t}
\end{align}
along with its average over noise realizations, $\psi(\br,\bp, t) \equiv \langle \hat{\psi}(\br,\bp, t)\rangle$. Using It\=o calculus~\cite{dean_langevin_1996}, we convert Eqs.~\eqref{eq:langevin-ud-pbp-r}-\eqref{eq:langevin-ud-pbp} to:
\begin{align} 
    \dot{\psi}&(\br,\bp,t) =  -\nabla \cdot \Big[\bp \psi\Big] + \nabla_\bp \cdot \bigg[\gamma T(\br)\nabla_\bp \psi + \gamma \bp \psi \nonumber\\
    &\quad + \int d\br' d \bp' \nabla \U(\br-\br') \langle \hat{\psi}(\br,\bp) \hat{\psi}(\br',\bp')\rangle \bigg] \label{eq:kramers-ud-pbp}\;.
\end{align}
We note that, unlike in the overdamped case, there is no ambiguity about where $T(\br)$ enters in Eq.~\eqref{eq:kramers-ud-pbp} since it commutes with $\nabla_\bp$. The overdamped ($\gamma\rightarrow \infty$) limit of these dynamics corresponds to the It\=o time-discretization of {OBP}s~\cite{van_kampen_relative_1988}{, as long as the friction $\gamma$ is spatially uniform. We study this overdamped system} in Sec.~\ref{sec:od-pbps} along with more general discretization schemes.

As we show below, non-uniform temperatures never lead to {directed currents for UBP}s, both in the non-interacting and interacting cases. We first show this numerically in Sec.~\ref{sec:ud-pbp-phenom}. We rule out TRS as a possible explanation in Sec.~\ref{sec:ud-pbp-trs}, since {UBP}s with inhomogeneous temperatures are irreversible both in the presence and in the absence of interactions. Finally, we show in Sec.~\ref{sec:ud-pbp-mom-cons-eos} that the lack of {directed} currents is due to an effective bulk momentum conservation in the steady state that arises from the existence of a stress tensor.

\subsection{Numerics and phenomenology}\label{sec:ud-pbp-phenom}

\begin{figure*}
    \begin{tikzpicture}
    \path (-0,0) node {\includegraphics{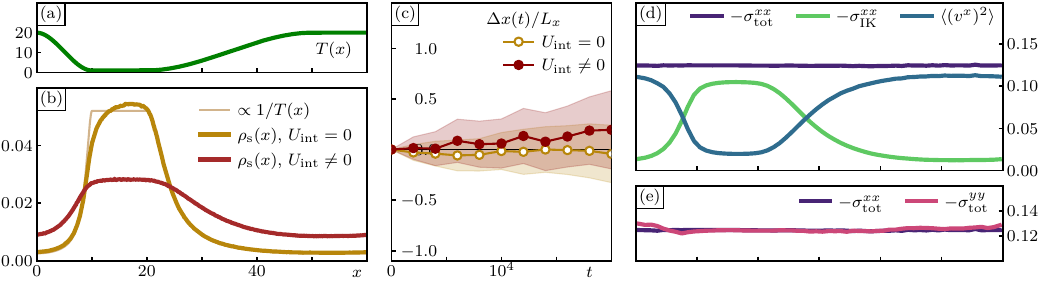}};
    \end{tikzpicture}
    \caption{\textbf{{Directed} currents are prevented in a two-dimensional system of {UBP}s by an emergent bulk momentum conservation.} {\bf (a)} The temperature landscape $T(x)$ used in simulations. {\bf (b)} The steady-state density distribution of {UBP}s both without (gold) and with  (dark red) interactions. The density in the $\gamma\rightarrow \infty$ limit, $\propto 1/T(x)$, is shown for reference (thin tan line). {\bf (c)} Directly measuring the net particle displacement over time, both without and with interactions, reveals the absence of {directed} current.  {\bf (d)} In the steady state, variations of the momentum flux (blue) are balanced by those of the Irving-Kirkwood stress (green), resulting in a homogeneous pressure $\sigma_{\rm tot}^{xx}$ and preventing the emergence of a steady current. {\bf (e)} The $(y,y)$ component of the stress tensor, $\sigma_{\rm tot}^{yy}$, varies along $x$, unlike $\sigma_{\rm tot}^{xx}$. It is, however, constant along $y$ by symmetry, thus generating no current.
    Simulation details are provided in Appendix~\ref{appendix:sim-ud-pbp}.}
    \label{fig:ud-pbp-eos}
\end{figure*}

We report in Fig.~\ref{fig:ud-pbp-eos} the results of particle-based simulations of {UBP}s in two space dimensions in the temperature landscape shown in Fig.~\ref{fig:ud-pbp-eos}(a). We consider systems both with and without pairwise repulsive interactions. 
The qualitative difference with systems in thermal equilibrium can be seen in the steady-state density shown in Fig~\ref{fig:ud-pbp-eos}(b): In the presence of a temperature field $T(\br)$, the damping $\gamma$ plays a role {in setting the steady-state distribution} since the density profile of non-interacting {UBP}s differs from its overdamped limit $\rho(\br)\propto 1/T(\br)$. 
Yet, Fig.~\ref{fig:ud-pbp-eos}(c) shows that no current is observed for {UBP}s, even in the presence of interactions. 

\subsection{Time-reversal Symmetry}\label{sec:ud-pbp-trs}
For {UBP}s, TRS can be immediately ruled out as an explanation for the lack of steady currents. Indeed, {UBP}s already exhibit a non-zero entropy production rate (EPR) in the non-interacting limit with spatially varying $T(\br)$~\cite{celani_anomalous_2012}.  To compute the EPR in the interacting case, we start from the definition:
\begin{align}
    \sigma &= \lim_{t_\mathrm{f}\rightarrow \infty} \frac{1}{t_\mathrm{f}} \left\langle \ln \left[\frac{\mathbb{P}\big[\{\br_i(t)\}\big|\{\br_i(0),\dot\br_i(0)\}\big]}{\mathbb{P}\big[\{\br^R_i(t)\}\big|\{\br^R_i(0),\dot\br^R_i(0)\}\big]} \right] \right\rangle \;.
    \label{eq:ud-pbp-sigma-both-dof}
\end{align}
where $\{\br^R_i(t)\}$ is the time-reverse of $\{\br_i(t)\}$, defined as
\begin{equation}
    \br_i^R(t) = \br_i(t_\mathrm{f}-t)\;.
\end{equation}
{We note that, throughout our work, we only use $\sigma$ as defined in Eq.~\eqref{eq:ud-pbp-sigma-both-dof} to detect whether the occurrence of trajectories $\{\br_i(t)\}$ and $\{\br_i^R(t)\}$ can be statistically distinguished. When $\sigma=0$, this is not the case and {directed} currents cannot emerge. 
We do not attempt to connect $\sigma$ to thermodynamics, for instance to estimate the heat dissipated in the system~\cite{celani_anomalous_2012}. This would require making sure that the models we study are thermodynamically consistent~\cite{fodor2022irreversibility}, for instance by specifying how non-uniform fluctuation sources are created and maintained.}

In Appendix~\ref{appendix:path-integ-upbp-int}, we show how to compute $\sigma$ using a Stratonovich time discretization of Eqs.~\eqref{eq:langevin-ud-pbp-r}-\eqref{eq:langevin-ud-pbp}. 
One can then define a fluctuating EPR density field $\hat{\sigma}(\br)$ such that $\int d\br \langle \hat{\sigma}(\br)\rangle=\sigma$, which reads:
\begin{align}
    \hat{\sigma}(\br) &= \sum_{i} \frac{\delta(\br-\br_i)}{T(\br_i)}  \dot{\br}_i \cdot \Big[ \sum_j \bfint(\br_i-\br_j) - \ddot{\br}_i \Big]\;.
    \label{eq:ud-pbp-epr-density}
\end{align}
The EPR is non-vanishing even in the non-interacting limit, $\bfint=0$, in agreement with~\cite{celani_anomalous_2012}. Equation~\eqref{eq:ud-pbp-epr-density} reveals that interactions introduce new mechanisms for entropy production. As we show below, interaction-induced irreversibility will prove generic in all systems considered in this article. 

{We conclude that UBPs in asymmetric temperature landscape break both TRS and parity symmetry, and yet lack steady directed currents. They are thus ``genuine'' exceptions to the ratchet principle, since the latter states that breaking all forms of TRS and parity symmetries should generically lead to a directed current~\cite{denisov_tunable_2014}. As we show below, an effective conservation law for the momentum of UBPs explains the absence of currents. }

\subsection{Bulk steady-state momentum conservation}\label{sec:ud-pbp-mom-cons-eos}
To understand the lack of {directed} currents in {UBP}s, we marginalize Eq.~\eqref{eq:kramers-ud-pbp} over momentum variables. To do so, we define the fluctuating density $\hat{\rho}$, momentum $\hat{\bP}$, and `nematic' order $\hat{\bQ}$ fields as
\begin{align}
    \label{eq:ud-pbp-rho-def}\hat{\rho}(\br) &\equiv \sum_i \delta(\br-\br_i) = \int d \bp\, \hat{\psi}(\br,\bp) \, ,\\
    \label{eq:ud-pbp-J-def}\hat{\bP}(\br) &\equiv \sum_i \bp_i \delta(\br-\br_i) = \int d \bp \, \hat{\psi}(\br,\bp) \bp \, ,\\
    \label{eq:ud-pbp-Q-def}\hat{\bQ}(\br) &\equiv \sum_i \Big(\bp_i \otimes \bp_i - \mathbb{I} \,T(\br_i) \Big) \delta(\br-\br_i) \\
    &= \int d \bp \,\hat{\psi}(\br,\bp) \big(\bp \otimes \bp - \mathbb{I} T(\br) \big)\;.\nonumber
\end{align}
We denote their averages as $\rho = \langle \hat{\rho}\rangle$, $\bP = \langle \hat{\bP}\rangle$, and $\bQ = \langle \hat{\bQ}\rangle$, respectively.

Integrating Eq.~\eqref{eq:kramers-ud-pbp} over $\bp$ then leads to
\begin{equation}
    \dot{\rho} = -\nabla \cdot \bP\label{eq:ud-pbp-rhodot}\;,
\end{equation}
which shows that, since the particle mass has been taken to unity, momentum density and particles flux coincide. Multiplying Eq.~\eqref{eq:kramers-ud-pbp} by $\bp$ and integrating over  $\bp$ then leads to
\begin{align}
    \dot{\bP} &= -\nabla \cdot \left[\bQ + \mathbb{I} \rho T\right] \nonumber\\
    &\quad- \int d\br' \nabla \U(\br-\br') \langle \hat{\rho}(\br)\hat{\rho}(\br')\rangle - \gamma \bP\;.\label{eq:ud-pbp-J}
\end{align}
In the steady state, $\dot{\bf P} = 0$, and, using $\bJ=\bP$, we find that
\begin{align}
    \bJ &= - \nabla \cdot \frac{\bQ + \mathbb{I} \rho T - \bsigma_\subIK}{\gamma} \equiv \nabla \cdot \frac{\bsigma_{{\rm tot}}}\gamma\;,\label{eq:ud-pbp-J-eos}
\end{align}
where we have absorbed the interaction term into the Irving-Kirkwood stress tensor that satisfies~\cite{irving1950statistical}
\begin{equation}\label{eq:sigmaIK}
\nabla \cdot \bsigma_\subIK = - \int d\br' \nabla \U(\br-\br') \langle \hat \rho(\br)\hat\rho(\br')\rangle\;.
\end{equation}
{For completeness, a derivation of $\bsigma_\subIK$ is reported in Appendix~\ref{app:irving-kirkwood}.} Using periodic boundary conditions, {Eq.~\eqref{eq:ud-pbp-J-eos} implies that the modulation of temperature along $\hat x$ does not induce a steady {directed} current $\bJ=J\hat x$ since}
\begin{equation}\label{eq:intJ}
  {J \hat x=\frac 1 {L^d}\int d\br \, \bJ(\br) \propto \int d\br\,\nabla \cdot \bsigma_{\rm tot}=0\;.}
\end{equation} 
{The momentum-conserving nature of the temperature inhomogeneities, which leads to $\bJ \propto \nabla \cdot [\bQ + \mathbb{I}\rho T]$ in Eq.~\eqref{eq:ud-pbp-J-eos}, thus enforces the lack of {directed} currents through Eq.~\eqref{eq:intJ}.}

\subsection{Summary}
While {UBP}s in spatially-varying temperature fields lack TRS, a {directed} current is prevented by the existence of a generalized stress tensor that enforces bulk momentum conservation. This makes {UBP}s a genuine exception to the ratchet principle. 

\section{Overdamped Passive Brownian Particles}\label{sec:od-pbps}

Let us now consider overdamped {Passive Brownian Particles (OBPs)} in a temperature field. In $d$ space dimensions, their dynamics read
\begin{equation}
    \dot{\br}_i (t) \overset{\alpha}{=} -\sum_{j=1}^N \nabla \U(\br_i-\br_j) + \sqrt{2T(\br_i)} \bfeta_i (t) \;,
    \label{eq:od-pbp-alpha-langevin}
\end{equation}
where we have taken the mobility to be unity. The symbol $\overset{\alpha}{=}$ reminds us that Eq.~\eqref{eq:od-pbp-alpha-langevin} is only well defined once its time discretization is specified. Introducing a time step $\Delta t$ and a discretized time $t_k=k\Delta t$, the displacement $\Delta \br_i^k=\br_i^{k+1}-\br_i^k$ of particle $i$ between times $t_k$ and $t_{k+1}$ is then given by:
\begin{equation}
    \Delta \br_i^k = -\Delta t\sum_{j=1}^N \nabla \U(\br_i^{k,\alpha}-\br_j^{k,\alpha}) + \sqrt{2 T(\br_i^{k,\alpha})} \Delta \bfeta_i^k\;,
    \label{eq:od-pbp-alpha-langevin-disc}
\end{equation}
where the $\{\Delta \bfeta_{i}^k\}$ are independent centered Gaussian random noises such that $\langle \Delta \eta_{i,\mu}^k \Delta \eta_{j, \nu}^\ell \rangle = \delta_{ij} \delta_{\mu \nu} \delta_{k \ell} \Delta t$. In Eq.~\eqref{eq:od-pbp-alpha-langevin-disc},
\begin{equation}
    \br_i^{k,\alpha}=\alpha \br_i^{k+1}+(1-\alpha) \br_i^{k}=\br_i^{k} + \alpha \Delta \br_i^k\;.
\end{equation}

Different values of $\alpha$ correspond to different physical processes. Note that the large damping {($\gamma \to \infty$)} limit of the underdamped colloids studied in Sec.~\ref{sec:ud-pbps} leads to the It\=o discretization $\alpha=0${, as long as $\gamma$ is spatially uniform~\cite{van_kampen_relative_1988}}.  Other physical processes, that cannot be interpreted as Brownian colloids with position-dependent temperature, may lead to Eq.~\eqref{eq:od-pbp-alpha-langevin} with other discretizations~\cite{van_kampen_diffusion_1988}. For the sake of generality, we thus consider the case of a generic $\alpha$ discretization. For simplicity, we still refer to the amplitude of the multiplicative noise as a `temperature', even when $\alpha\neq 0$.

\begin{figure*}
    \centering
\begin{tikzpicture}
    \path (-0,0) node {\includegraphics{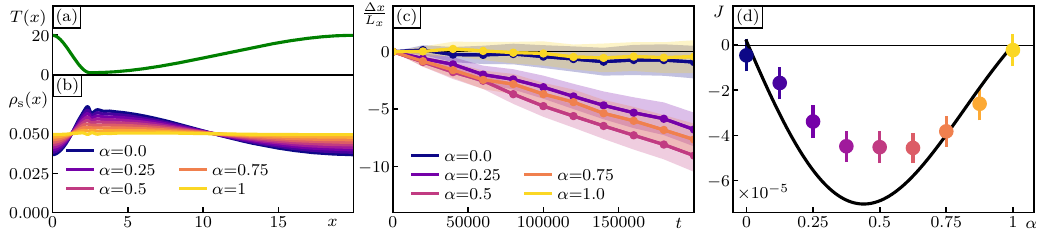}};
\end{tikzpicture}
    \caption{\textbf{Interaction-induced current for {OBP}s with different time discretizations}. {\bf (a)} Particles interacting via a soft repulsive harmonic potential, Eq.~\eqref{eq:appendix-Uint}, are placed in a temperature landscape $T(x)$. {\bf (b)} Steady-state particle density as the time-discretization parameter $\alpha$ is varied. {\bf (c)} The net integrated displacement of the particles show the emergence of nonzero currents for discretizations $0<\alpha<1$. {\bf (d)} Steady-state particle current as $\alpha$ is varied. Momentum conservation prevents a {directed} current in the It\=o-discretized ($\alpha$=0) system. The H\"anggi-discretized ($\alpha$=1) system is equivalent to a time-reversed It\=o dynamics and thus also current free. Finally, no such conservation exists in other discretizations which thus lead to {directed} currents. The mean-field prediction for the current $J$, to 1st-order in the interaction strength, Eq.~\eqref{eq:od-pbp-alpha-J1}, is plotted in black. Simulation details are provided in Appendix~\ref{appendix:sim-od-pbp}.}
    \label{fig:od-pbp-disc-induced-current}
\end{figure*}

These microscopic dynamics translate into an evolution equation for the fluctuating empirical density field $\hat{\rho}$, defined as:
\begin{align}
    \hat{\rho}(\br,t) &\equiv \sum_i \delta(\br-\br_i(t))\;,
\end{align}
and for its average $\rho(\br)\equiv \langle \hat{\rho}(\br)\rangle$. For an $\alpha$ discretization, the evolution of $\rho$ reads~\cite{de_pirey_path_2022}
\begin{align}
    \dot \rho(\br,t) &= \nabla \cdot \bigg[T(\br)^\alpha \nabla \Big\{\rho(\br,t) T(\br)^{1-\alpha}\Big\} \label{eq:od-pbp-alpha-fpe}\\
    &\qquad + \int d \br' \big\langle\hat{\rho}(\br,t)\hat{\rho}(\br',t) \big\rangle \nabla \U(\br-\br')\bigg]\nonumber\\
    &\equiv -\nabla \cdot \bJ(\br,t)\;\,
\end{align}
where we have introduced the particle current $\bJ(\br,t)$.

Below, we first review in Sec.~\ref{sec:od-pbp-numerics-phenom} the phenomenology of {OBP}s using numerical simulations. We then explore their TRS in Sec.~\ref{sec:od-pbp-trs}. We first show that, for all $\alpha$, non-interacting {OBP}s do not display steady currents due to an emergent TRS  {in position space}. We then calculate the non-zero entropy production induced by interactions between the particles. 
In Sec.~\ref{sec:od-pbp-eos}, we show that effective momentum conservation forbids interaction-induced currents for the It\=o-discretized system and the `H\"anggi' ($\alpha=1$) time discretization, which is the time reverse of an It\=o-discretized system. On the contrary, for $0<\alpha<1$, the multiplicative noise leads to net momentum sources and the emergence of {directed} currents. 
Finally, in Sec.~\ref{sec:od-pbp-whatever}, we characterize the onset of this {directed} current perturbatively in the interaction strength between particles.

\subsection{Numerics and phenomenology}\label{sec:od-pbp-numerics-phenom}
In the absence of interactions ($\U=0$), the steady-state density distribution satisfies
\begin{align}
    \rho_{\rm s}(\br) &\propto T(\br)^{\alpha-1}\;,
    \label{eq:alpha-disc-ss-rho}
\end{align}
as can be inferred from Eq.~\eqref{eq:od-pbp-alpha-fpe}. The resulting current $\bJ(\br)$ then vanishes, making this system an apparent exception to the ratchet principle.

Again, it is interesting that adding interactions induces a current for these systems \textit{only} for discretizations  $0 < \alpha < 1$, as demonstrated in Fig.~\ref{fig:od-pbp-disc-induced-current}. The $\alpha=0$ (It\=o) and $\alpha=1$ (H\"anggi) discretizations do not admit interaction-induced currents.

\subsection{Time-reversal symmetry}\label{sec:od-pbp-trs}
We show below that noninteracting {OBPs obey TRS in position space} regardless of their discretization. We prove this first by computing the path probabilities in Sec.~\ref{sec:od-pbp-trs-ni} and then by demonstrating the symmetry of the Fokker-Planck operator in Sec.~\ref{sec:od-pbp-trs-operator}. We then turn to interacting particles in Sec.~\ref{sec:od-pbp-trs-int}. In the rest of this article, we use either the path-probability or operator approach, depending on the context. {We stress again that TRS only means here that forward and backward trajectories cannot be statistically distinguished.}

\subsubsection{TRS at the trajectory level}\label{sec:od-pbp-trs-ni}

First, we set $\U=0$ to consider noninteracting particles. The system then reduces to a single-body problem and we omit below the particle index. When calculating the probabilities of trajectories, the multiplicative noise $T(\br)$ calls for some care~\cite{de_pirey_path_2022}. The corresponding time-discretized calculations are detailed in Appendix~\ref{appendix:path-integ-pbp-ni}, leading to a path probability for a trajectory ${\br} (t)$ starting at a given position $\br(0)$:
\begin{align}\label{eq:PINI}
    \mathbb{P}&[\br(t)|\br(0)] \propto F(\{\overline{\br}^k\})\\
    &\quad \times\exp\bigg[ \int_0^{\tf} dt \bigg\{-\frac{\big|\dot{\br} - \big(\alpha-\frac12\big) \nabla T(\br)\big|^2}{4 T(\br)}\nonumber\\
    &\quad +\frac{\left(\alpha-\frac12\right) }{2} \left( \frac{|\nabla T({\br})|^2}{2T({\br})} - \nabla^2 T({\br})\right) - \frac{\dot{\br} \cdot \nabla T({\br})}{4T({\br})}\bigg\}\bigg]\;,\nonumber
\end{align}
where $F$ is a function of the discretized positions at the midpoint of each timestep, $\overline{\br}^k$, defined as
\begin{equation}
    \overline{\br}^k=\frac 1 2 {\br}^k + \frac 1 2 {\br}^{k+1}\;.
\end{equation}
While Eq.~\eqref{eq:PINI} describes the path-probability of Langevin dynamics for any values of $\alpha$, the time-discretization used to construct \textit{the path-integral} is the Stratonovich one, so that the standard chain rule can be used to compute integrals appearing in the exponent~\cite{de_pirey_path_2022}.

The probability of the reverse trajectory is found by replacing the trajectories with their time reverses,
\begin{align}
    \br(t)&\mapsto \br^R(t)= \br(\tf-t)\;,\\
    \dot{\br}(t)&\mapsto \dot{\br}^R(t)=-\dot{\br}(\tf-t)\;.
\end{align}
Taking the ratio of the forward and reverse path probabilities results in the cancellation of all terms symmetric under time reversal, such as $|\dot{\br}|^2$, $F(\{\bar\br_k\})$, and all terms that depend solely on $\br$. One then finds:
\begin{align}
    \frac{\mathbb{P}[\br(t)|\br(0)]}{\mathbb{P}[\br^R(t)|\br^R(0)]} &= \exp \left\{\int_0^{\tf} dt\Big(\alpha-1\Big) \frac{\dot{\br}\cdot \nabla T(\br)}{T(\br)}\right\}\nonumber\\
    &= \exp\left\{\Big(\alpha-1\Big)\ln \left[\frac{T(\br(\tf))}{T(\br(0))}\right]\right\}\nonumber\\
    &= \frac{T(\br(\tf))^{\alpha-1}} { T(\br(0))^{\alpha-1}} = \frac{\rho_{\rm s}[\br(\tf)]} {\rho_{\rm s}[\br(0)]}\;.
\end{align}
One thus has that the probability of observing forward and backward trajectories are equal: $\mathbb{P}[\br(t)|\br(0)]\rho_{\rm s}[\br(0)]=\mathbb{P}[\br^R(t)|\br^R(0)]\rho_{\rm s}[\br^R(0)]$. {Noninteracting OBPs} with inhomogeneous temperature and arbitrary discretization are thus time-reversal symmetric, despite having a non-Boltzmann steady-state density. This explains why they do not lead to {directed} currents.

\subsubsection{TRS at the operator level}\label{sec:od-pbp-trs-operator}

Time-reversal symmetry can also be read in the existence of a conjugation relation between the evolution operator and its adjoint~\cite{risken1972solutions,risken1996fokker,gardiner2009stochastic}. While less physically transparent, the calculation is actually easier in the non-interacting case, and nicely complements the path integral approach presented above. It is also a gentle introduction to the active case, that will prove more involved.
For a single {OBP} in a temperature field, the Fokker-Planck operator ${\cal L}_{\rm FP}$, which satisfies ${\dot \rho} = - {\cal L}_{\rm FP} \rho$, is given by ${\cal L}_{\rm FP}=-\nabla \cdot T^{\alpha} \nabla T^{1-\alpha}$. Its adjoint is ${\cal L}_{\rm FP}^\dag = -T^{1-\alpha} \nabla \cdot T^\alpha \nabla$. Direct calculation then shows that:
\begin{align}
    \nonumber \rho_{\rm s} {\cal L}_{\rm FP}^\dag \rho_{\rm s}^{-1} &= -\frac{C}{T^{1-\alpha}} T^{1-\alpha} \nabla \cdot T^\alpha \nabla  \frac{T^{1-\alpha}}{C}\\
    &= -\nabla \cdot T^\alpha \nabla T^{1-\alpha}  = {\cal L}_{\rm FP}\;,
\end{align}
which implies detailed balance with respect to the non-Boltzmann steady-state distribution $\rho_{\rm s}$.

\subsubsection{Interaction-induced TRS violation}\label{sec:od-pbp-trs-int}
\begin{figure}
\begin{tikzpicture}
    \path (0,0) node {\includegraphics{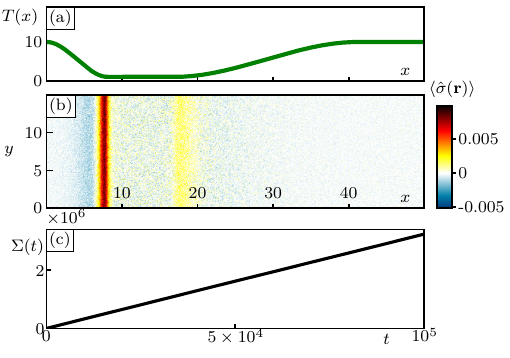}};
\end{tikzpicture}
    \caption{\textbf{EPR measured in simulations of {OBP}s}. {\bf (a)} Temperature
      landscape. {\bf (b)} Heat map of
      the average EPR density field $\langle
      \hat{\sigma}(\br)\rangle$. {\bf (c)} Net entropy production up to
      time $t$. Adapted from Fig.~2 of our companion letter~\cite{metzger_exceptions_letter}. Simulation details are provided in Appendix~\ref{appendix:sim-od-pbp}.}
    \label{fig:EOSpassive}
\end{figure}

We now extend the path-integral calculation of Sec.~\ref{sec:od-pbp-trs-ni} to the interacting case, $\U\neq 0$, using again the general $\alpha$ discretization~\eqref{eq:od-pbp-alpha-langevin-disc}. This calculation is detailed in Appendix~\ref{appendix:path-integ-pbp-int} and leads to:
\begin{align}
    \mathbb{P}&[\{\br_i(t)\}|\{\br_i(0)\}] = K(\{\overline{\br}^k_i,|\Delta{\br}^k_i|^2\})\\
    &\times\exp\Bigg[- \int_0^{\tf} dt \sum_{i=1}^N \Bigg\{\frac{\dot{\br}_i \cdot \nabla T(\br_i)}{4T(\br_i)}\nonumber\\
    &\quad -\frac{\dot{\br}_i \cdot \left[\big(\alpha-\frac12\big) \nabla T(\br^i) + \sum_j \mathbf{f}_{\rm int}(\br_i-\br_j)\right]}{2 T(\br_i)} \Bigg\}\Bigg] \nonumber\;.
\end{align}
Again, we can use this expression to evaluate the probability to observe the reverse trajectory $\br^R_i(t)= \br_i(\tf-t)$, taking care of the discretization issues. We then calculate the entropy production as
\begin{align} \label{eq:PBP_int_EP}
    \Sigma &\equiv \ln \left[\frac{\mathbb{P}[\{\br_i(t)\}|\{\br_i(0)\}]}{\mathbb{P}[\{\br_i^R(t)\}|\{\br_i^R(0)\}]}\right] \\
    &= -\int_0^{\tf} dt \sum_{i=1}^N \frac{\dot{\br}_i \cdot \big[ (\alpha-1)\nabla T(\br_i) - \sum_j \mathbf{f}_{\rm int}(\br_i-\br_j)\big]}{T(\br_i)} \nonumber\\
    &= \sum_{i=1}^N \ln \left[\frac{T(\br_i(\tf))^{\alpha-1}}{T(\br_i(0))^{\alpha-1}}\right]
    + \int_0^{\tf} dt \sum_{i,j=1}^N \frac{\dot{\br}_i \cdot \bfint(\br_i-\br_j)}{T(\br_i)}.\nonumber
\end{align}
The boundary terms in Eq.~\eqref{eq:PBP_int_EP} do not contribute to the EPR and we find:
\begin{equation}
    \sigma = \lim_{\tf \to \infty} \frac{ 1}{\tf} \Sigma = \int d\br \langle \hat \sigma (\br) \rangle\;,
\end{equation}
where in the last step we have invoked ergodicity and introduced an entropy-production rate density
\begin{align}
    \hat{\sigma}(\br) &= -\sum_{i,j} \delta(\br-\br_i) \frac{\dot{\br}_i \cdot \nabla \U(\br_i-\br_j)}{T(\br_i)}\;.
    \label{eq:pbp-epr-density}
\end{align}
Interestingly, this expression is independent of the discretization $\alpha$ and shows that $\sigma$ is generically non-zero. The irreversibility of the dynamics is confirmed using particle-based simulations with an It\=o discretization ($\alpha=0$) in Fig.~\ref{fig:EOSpassive}.

\subsection{Effective momentum conservation}\label{sec:od-pbp-eos}

In this section, we show that, despite having a finite EPR,  {OBP}s with It\=o discretization do not exhibit {directed} currents. As for {UBP}s, this can be explained by an emergent bulk momentum conservation which prevents the existence of the momentum sources required to rectify fluctuations into motion.  We first rewrite Eq.~\eqref{eq:od-pbp-alpha-fpe} as
\begin{align}
    \nonumber \dot{\rho} &= \nabla \cdot \bigg\{  \nabla (T\rho)  - \int d\br' \langle \hat{\rho}(\br)\hat{\rho}(\br')\rangle \bfint(\br-\br') - \alpha \rho \nabla T \bigg\}\nonumber\\
    &\equiv \nabla \cdot \big\{\nabla \cdot \left[\mathbb{I} \rho T - \bsigma_\subIK\right] - \alpha \rho \nabla T \big\}\nonumber\\
    &\equiv -\nabla \cdot \big\{\nabla \cdot \bsigma_{\rm tot} + \delta \bF_{\rm P}^{(\alpha)}\big\}\;, \label{eq:pbp_int_stress_eom}
\end{align}
where we have again introduced the Irving-Kirkwood stress tensor defined in Eq.~\eqref{eq:sigmaIK}. The term that could not be written as the divergence of a stress tensor leads to the nonconservative force $\delta \bF_{\rm P}^{(\alpha)} \equiv \alpha \rho \nabla T$.
Thus we can write the current as the sum of momentum-conserving and nonconserving parts as
\begin{align}
    \bJ &= \nabla \cdot \bsigma_{\rm tot} + \delta \bF_{\rm P}^{(\alpha)}. 
    \label{eq:od-pbp-J-sig-mom-source}
\end{align}
This has important, discretization-dependent consequences for the presence of steady-state {directed} currents.

First, let us consider the case of It\=o discretization $\alpha=0$. In this case, the nonconservative force vanishes $\delta \bF_{\rm P}^{(\alpha)}=0$, and the current can be written as the divergence of a stress tensor, as for {UBP}s (Sec.~\ref{sec:ud-pbp-mom-cons-eos}). {Thus, modulations of temperature along $\hat x$ cannot induce a steady {directed} current $\bJ=J \hat x$ since $L^d J = \int d\br \, \bJ=0$ by the divergence theorem. There is no {directed} current in the steady state for $\alpha=0$.}

When $\alpha\neq 0$, however, $\delta \bF_{\rm P}^{(\alpha)}$ may act as a momentum source in regions with temperature gradients. Thus there is no apparent symmetry preventing the emergence of {directed} currents. 

\subsection{Emergence of interaction-induced current}
\label{sec:od-pbp-whatever}
In this section, we demonstrate that a current arises for $0<\alpha<1$ using a factorization approximation. To make progress, we focus on systems with translational symmetry in all but one directions, using $T(\br)=T(x)$. Further, to close Eq.~\eqref{eq:pbp_int_stress_eom}, we factorize the pair-correlation function as
\begin{align}
    \langle \hat{\rho}(x) \hat{\rho}(x')\rangle \approx \langle \hat{\rho}(x)\rangle \langle \hat{\rho}(x')\rangle = \rho(x) \rho(x')\;.\label{eq:rho-mf-fact-od-pbp}
\end{align}
This approximation becomes accurate when correlations between particles can be neglected, e.g. in the weak interaction, large density, or high-dimensional limits. These simplifications allow us to rewrite Eq.~\eqref{eq:od-pbp-alpha-fpe} as
\begin{align}
    \dot{\rho} &= \partial_x \Big[T^\alpha \partial_x \left(\rho T^{1-\alpha}\right) + \rho(x) \int dx' \rho(x') \U'(x-x')\Big]\nonumber \\
    &= \partial_x \Big[\partial_x \big( \rho T\big) - \alpha \rho \partial_x T + \rho \partial_x V_{\rm eff}\Big]\nonumber\\
    &= \partial_x \Big[\partial_x \big(\rho T\big) - \rho F_\eff(x)\Big]\;,\label{eq:mf-od-pbp-1d}
\end{align}
where we have defined an effective potential $V_\eff$ and effective total force $F_\eff$ as:
\begin{align}
    V_\eff(x) &\equiv (\U\ast\rho)(x) = \int dx' \rho(x') \U(x-x')\label{eq:Veff-convolution-od-pbp}\;,\\
    F_\eff(x)&\equiv \alpha T'(x) - \partial_x V_\eff(x)\;.\label{eq:Feff-convo}
\end{align}
Equations~\eqref{eq:mf-od-pbp-1d}-\eqref{eq:Feff-convo} show that the factorization approximation maps the system onto the dynamics of an $\alpha$-discretized single particle that experiences a mean-field effective potential $V_\eff$ created by the other particles, or equivalently of an It\=o-discretized particle that experiences an effective force field $F_{\rm eff}$.

Before solving this nonlinear equation perturbatively (and self-consistently) in the interaction strength, we first review the simpler, non-interacting problem where $F_\eff(x)$ is an arbitrary \textit{external} force, which was solved by van Kampen~\cite{van_kampen_relative_1988}. In addition to its pedagogical purpose, this allows us to fix a typo in~\cite{van_kampen_relative_1988} and introduce useful notations.

\subsubsection{A single particle in potential, temperature, and mobility fields}\label{sec:van-kampen-review-1d-fokker-planck}
Van Kampen considered a single {OBP} in a temperature field $T_\eff(x)$, mobility field $\mu_\eff(x)$, and external force field $F_\eff(x)$, described by the Fokker-Planck equation
\begin{align}
    \dot{\rho} &= \partial_x \left\{\mu_\eff \left[\partial_x \big(T_\eff\rho\big) - \rho F_\eff\right]\right\} \equiv -\partial_x J\;.\label{eq:fpe-1d-pbp-van-kampen}
\end{align}
Direct algebra shows that the steady-state solution is given by~\cite{van_kampen_relative_1988}
\begin{align}
    &\rho(x) = \frac{e^{-\Phi(x)}}{T_\eff(x)}\bigg[\rho(0) T_\eff(0) - J \int_0^x du \frac{e^{\Phi(u)}}{\mu_\eff(u)}\bigg]\;,\label{eq:VKsol}\\
    &J = \frac{e^{-\Phi(L)}-1}{\int_0^L \frac{dx e^{-\Phi(x)}}{T_\eff(x)} \left[\int_0^x \frac{du e^{\Phi(u)}}{\mu_\eff(u)} + \int_x^L \frac{du e^{\Phi(u)-\Phi(L)}}{\mu_\eff(u)} \right]}\;,\label{eq:VKsolJ}
\end{align}
where $\rho(0)$ is a constant such that $\rho$ is normalized~\cite{van_kampen_relative_1988} and the ``pseudo-potential" $\Phi$ is given by
\begin{align}
    \Phi(x) &\equiv -\int_0^x \frac{F_\eff(x')}{T_\eff(x')} dx'\;.\label{eq:od-pbp-van-kampen-phi-def}
\end{align}
We note that $T_\eff(0)$ in Eq.~\eqref{eq:VKsol} was mistakenly written as $T_\eff(x)$ in~\cite{van_kampen_relative_1988}.

 One sees that a nonzero current arises if and only if $\Phi(L) \neq 0$; i.e., when $\Phi$ is aperiodic:
\begin{align}
    J \neq 0 \quad \Longleftrightarrow \quad \Phi(L) \neq 0\;.\label{eq:J-nonzero-condition-Phi-aperiodic}
\end{align}
Let us now show how we can use the non-interacting solution given in Eqs.~\eqref{eq:VKsol}-\eqref{eq:od-pbp-van-kampen-phi-def} to solve the interacting problem perturbatively.

\if{
We will return to this useful solution later, e.g. in Sec.~\ref{sec:1d-rtps}, where it will allow us to exactly determine the steady state for 1-dimensional RTPs in an activity and potential landscape, when we approximate the effect of interaction by particles with a density field as an effect exerted by a potential field. Next, we will return to interacting, $\alpha$-discretized {OBP}s, and use the solution of the Fokker-Planck equation to evaluate the {directed} current.
}\fi

\subsubsection{Perturbation theory}\label{sec:od-pbp-1d-fpe-solve-van-kampen-mf}
To perform a perturbative expansion in the interaction strength, we introduce a rescaled potential $\Tilde{U}_{\rm int}$ such that $\U=\varepsilon \Tilde{U}_{\rm int}$ and we use $\varepsilon$ as a small parameter. We then use the solution~\eqref{eq:VKsol}-\eqref{eq:od-pbp-van-kampen-phi-def} to Eq.~\eqref{eq:mf-od-pbp-1d} with $T_\eff=T(x)$, $\mu_\eff=1$, and $F_\eff=\alpha T' - \varepsilon (\Tilde{U}_{\rm int}\ast\rho)'$ to show that the steady-state density satisfies the self-consistent integral equation:
\begin{align}
    \rho &= \frac{\exp \left(-\varepsilon \int_0^x du \frac{(\Tilde{U}_{\rm int}\ast \rho)'}{T}\right)}{T^{1-\alpha}} \nonumber\\
    &\quad \times \bigg[\kappa - J \int_0^x du \frac{\exp \left(\varepsilon \int_0^u du' \frac{(\Tilde{U}_{\rm int}\ast \rho)'}{T}\right)}{T^\alpha} \bigg]\label{eq:od-pbp-1d-mf-nonlin-rho-sol}
\end{align}
where $\kappa \equiv \rho(0) T(0)^{1-\alpha}$ is a normalization constant such that $\int dx \rho(x)=N$. Next, we insert the expansion
\begin{align}
    \rho(x) &= \sum_{k=0}^\infty \varepsilon^k \rho_k(x),\qquad J = \sum_{k=0}^\infty \varepsilon^k J_k
\end{align}
into Eq.~\eqref{eq:od-pbp-1d-mf-nonlin-rho-sol}, which we solve order by order. At zeroth order, we recover the solution Eq.~\eqref{eq:alpha-disc-ss-rho} which describes noninteracting particles: $\rho_0=\kappa_0/T^{1-\alpha}$ and $J_0=0$, where $\kappa_0 = N \big[\int_0^L T(u)^{\alpha-1}\big]^{-1}$. Then, the first order correction to the density field and current are given by
\begin{align}
    \rho_1 &= \frac{1}{ T^{1-\alpha}} \left\{\kappa_1 - \int_0^x du \bigg[\kappa_0\frac{(\Tilde{U}_{\rm int}\ast \rho_0)'(u)}{T(u)} + \frac{J_1 }{ T(u)^\alpha} \bigg]\right\}\;.\label{eq:od-pbp-alpha-rho1}\\
    J_1 &= - \frac{\kappa_0^2 }{ \int_0^L \frac{dx}{T(x)^\alpha}} \int_0^L dx \int_0^L dx' \frac{\tilde{U}_{\rm int}'(x-x')}{T(x) T(x')^{1-\alpha}}\;,\label{eq:od-pbp-alpha-J1}
\end{align}
where $\kappa_1$ is such that $\int_0^L \rho_1(x)dx=0$ and $J_1$ enforces the periodicity of $\rho_1$. We thus see that a non-vanishing current emerges at first order in the interaction amplitude. Already, this result captures semi-quantitatively the trend shown in numerical simulations of interacting particles, as seen in Fig.~\ref{fig:od-pbp-disc-induced-current}(d). The perturbation expansion can then be solved systematically to higher order, in the spirit of what will be done below for one-dimensional RTPs in Sec.~\ref{sec:1d-rtps}.

Consistent with Sec.~\ref{sec:od-pbp-eos}, $J_1$ vanishes for the It\=o discretization $\alpha=0$, since $\tilde {U}_{\rm int}'(x-x')$ is antisymmetric under $x \leftrightarrow x'$, while the denominator in Eq.~\eqref{eq:od-pbp-alpha-J1} is symmetric for $\alpha=0$. 
Interestingly, $J_1$ also vanishes for $\alpha=1$, since $J_1\propto [\tilde {U}_{\rm int}(x+L)-\tilde {U}_{\rm int}(x)]=0$ in this case.
For other values of $\alpha$, however, $J_1$ is generally nonzero if $T$ is asymmetric, consistent with the numerical results shown in Fig.~\ref{fig:od-pbp-disc-induced-current}(d).

Note that one can show the current to vanish exactly for the $\alpha=1$ H\"anggi discretization{, and not merely at the level of the perturbative expansion presented above}. Although it appears to lack momentum conservation according to Eq.~\eqref{eq:pbp_int_stress_eom}, a H\"anggi-discretized trajectory is equivalent to the time-reverse of an It\=o-discretized one, with a force of the opposite sign. In fact, one can perform the same mapping between any $\alpha$-discretized trajectory and a $(1-\alpha)$-discretized one, since Eq.~\eqref{eq:od-pbp-alpha-langevin-disc} can be re-written as
\begin{align}
    \label{eq:alpha-1minusalpha-mapping}-\Delta \br_i^{R,k} &= -\sum_{j} \nabla \U(\br_i^{R,k+1} - \br_j^{R,k+1}) \Delta t\\
    &\quad + \sqrt{2 T(\br_i^{R,k} + (1- \alpha) \Delta \br_i^{R,k+1})} \Delta \bfeta_i^{R,k+1} \nonumber
\end{align}
where the reverse trajectory is defined as $\br_i^{R,k}\equiv \br_i^{N-k}$. Setting $\alpha=0$, we find that the trajectory $\br_i^R$ is identical to a H\"anggi-discretized {OBP} with interaction potential $-\U$.
Since the current of an It\=o-discretized trajectory must be zero regardless of the sign of its force due to the effective momentum conservation, the time-reverse of any It\=o-discretized trajectory must also have zero current. Therefore, the interaction-induced {directed} current vanishes for both $\alpha=0$ and $\alpha=1$.

\subsection{Summary}
{OBP}s with inhomogeneous temperature, perhaps the oldest ``exception" to the ratchet principle~\cite{van_kampen_diffusion_1988,van_kampen_relative_1988}, display a surprisingly rich phenomenology when interactions and various choices of time discretizations are considered.

We have seen that spatially varying temperature fields cannot alone power steady-state currents, regardless of discretization scheme, due to the TRS of the dynamics {in position space. This effective} TRS is broken when interparticle interactions are introduced, as reflected in the EPR shown in Eq.~\eqref{eq:pbp-epr-density}, which does not depend on the discretization. Despite this TRS violation, It\=o-discretized {OBP}s obey a bulk momentum conservation which prevents them from exhibiting a current. Furthermore, H\"anggi-discretized {OBP}s are equivalent to the time-reverse of It\=o-discretized {OBP}s with an opposite force, and thus also exhibit no {directed} current.

On the contrary, for discretizations $0<\alpha<1$---and notably for the Stratonovich discretization $\alpha=1/2$---there are momentum sources, as shown in Eq.~\eqref{eq:pbp_int_stress_eom}, which power steady-state {directed} currents. We have measured these currents in particle-based simulations (Fig.~\ref{fig:od-pbp-disc-induced-current}) and analytically approximated them in Eq.~\eqref{eq:od-pbp-alpha-J1} using a factorization approximation and a perturbative expansion.

\section{Active-Brownian and Run-and-Tumble Particles}\label{sec:abp-rtp}

\begin{figure*}
\begin{tikzpicture}
\path (0,0) node {\includegraphics{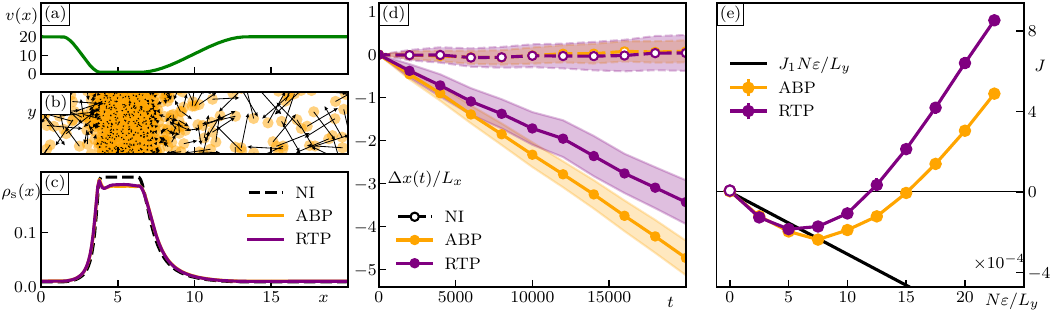}};
\end{tikzpicture}
    \caption{
    \textbf{Interaction-induced current for ABPs and RTPs in $d=2$ spatial dimensions.} We use a soft repulsive harmonic potential proportional to $\varepsilon$. {\bf (a)} Self-propulsion landscape $v(x)$ used in all simulations. Panels {\bf (b-d)} correspond to simulations with $N\varepsilon/L_y=7.5$. {\bf (b)} Snapshot of particle locations and self-propulsion vectors for an ABP simulation.  {\bf (c)} Steady-state density distributions, compared to the non-interacting result $\rho_{\rm s}(x)\propto 1/v(x)$. {\bf (d)} Net average displacement per particle over time for interacting (solid) and non-interacting (dashed) ABPs and RTPs. The shaded region indicates $\pm 3\sigma$, where $\sigma$ is the standard error of the mean. {\bf (e)} Average current vs. interaction strength, along with the mean-field prediction to first order in $\varepsilon$. Simulation details are provided in Appendix~\ref{appendix:sim-abp-rtp}.}
    \label{fig:abp-rtp-2d-phenomenology}
\end{figure*}

Next, we move our focus to Active Brownian Particles (ABPs) and Run-and-Tumble Particles (RTPs) with spatially varying activity. We consider these two systems together because, as we will demonstrate, their behaviors are quite similar to each other. Again, we introduce the models, briefly discuss their phenomenologies, and examine time-reversal symmetry and momentum conservation. Finally, we discuss the characterization of the emergence of interaction-induced currents using a perturbation theory.

We consider $N$ self-propelled particles in $d$ dimensions with the spatial dynamics
\begin{equation} \label{eq:ri_abp_rtp}
        \dot{\br}_i = v(\br_i) \bu_i(t) - \sum_j \nabla \U(\br_i-\br_j)\;,
\end{equation}
where $\br_i \in \mathbb{R}^d$ is the position of particle $i$, $\bu_i \in S^{d-1}$ is its orientation director on the $(d-1)$-sphere, and $\U$ is the pairwise interaction potential. For ABPs, the director undergoes rotational diffusion on the $(d-1)$-sphere, with diffusivity $D_r$. In $d=2$, which is the case where most of our computations will be done, $\bu_i(\theta_i)=(\cos\theta_i,\sin\theta_i)$ and the dynamics reads
\begin{equation}
    \dot \theta_i=\sqrt{2 D_r} \eta_i\;,
\end{equation}
where $\eta_i$ is a Gaussian white noise such that
\begin{equation}
    \langle \eta_i (t) \eta_j(t')\rangle = \delta_{ij} \delta (t-t')\;.
\end{equation}

For RTPs, the orientation director undergoes a complete randomization with a rate {$\tau^{-1}$}:
\begin{align} \label{eq:ui_rtp}
    \bu_i \stackrel{{\tau^{-1}}}{\rightarrow} \bu_i'
\end{align}
where $\bu_i'$ is sampled from a uniform distribution on $S^{d-1}$. For RTPs in 1d, $\bu_i$ is restricted to $\{+1,-1\}$. We denote the persistence time by $\tau=1/[(d-1)D_r]$ for ABPs.

We define the empirical density in $(\br,\bu)$ space as $\hat{\psi}(\br,\bu) = \sum_i \delta(\br-\br_i) \delta(\br-\bu_i)$, along with its average over noise realizations, $\psi(\br,\bu) = \langle \hat{\psi}(\br,\bu) \rangle$. Standard methods~\cite{tailleur_statistical_2008,dean_langevin_1996} allow us to convert~\eqref{eq:ri_abp_rtp}-\eqref{eq:ui_rtp} into:
\begin{align}
    \label{eq:fpe-abp-rtp}
    \dot{\psi}&(\br,\bu,t)= \mathcal{L}_\mathrm{r} \psi(\br,\bu)-\nabla \cdot \bigg[v(\br) \bu \psi(\br,\bu) \\
    &\quad - \int d\br' \int d\bu' \nabla \U(\br-\br') \langle \hat{\psi}(\br,\bu) \hat{\psi}(\br',\bu')\rangle\bigg]\;.\nonumber
\end{align}
We have defined an operator $\mathcal{L}_\mathrm{r}$ which accounts for the dynamics of the orientation $\bu$. For RTPs, it reads
\begin{align}
    \mathcal{L}_\mathrm{r} \psi &= -\frac{\psi(\br,\bu)}{{\tau}} + \frac{1}{{\tau}\Omega_{d}} \int d^{d-1} \bu' \psi(\br,\bu')
    \label{eq:rtp-orientation-dynamics}
\end{align}
while for ABPs it is given by
\begin{align}
    \mathcal{L}_\mathrm{r} \psi &= D_r \Delta_{\bu} \psi,
    \label{eq:abp-orientation-dynamics}
\end{align}
where $\Delta_{\bu}$ is the Laplacian on the $(d-1)$-dimensional sphere.

In the following sections, we show through a combination of exact calculations, motivated approximations, and particle-based simulations that non-interacting ABPs and RTPs in $d$ dimensions are prevented from exhibiting {directed} currents by a hidden TRS {in the position degrees of freedom}, which is destroyed in the presence of interactions. The lack of momentum conservation then generically induces a {directed} current, which we calculate pertubatively using a factorization approximation.

\subsection{Numerics and phenomenology}\label{sec:abp-rtp-phenom}
ABPs have been shown numerically to experience interaction-induced current in~\cite{stenhammar_light-induced_2016}. In Fig.~\ref{fig:abp-rtp-2d-phenomenology}, we characterize quantitatively the emergence of {directed} currents for ABPs and RTPs in $d=2$ spatial dimensions. Non-interacting ABPs and RTPs in an activity landscape $v(\br)$ relax to an isotropic, current-free steady state with $\psi_\mathrm{s}(\br,\bu)\propto 1/v(\br)$~\cite{schnitzer_theory_1993,cates_motility-induced_2015,arlt_painting_2018,frangipane_dynamic_2018}, as can be inferred from Eq.~\eqref{eq:fpe-abp-rtp}. Introducing interactions perturbs the density distribution and destroys the current-free state [Fig.~\ref{fig:abp-rtp-2d-phenomenology}(d)]. The resulting current can be very strong and sometimes exhibits reversals [Fig.~\ref{fig:abp-rtp-2d-phenomenology}(e)].

\subsection{Time-reversal symmetry}\label{sec:abp-rtp-trs}
In this section we show that, in the absence of interactions, ABPs and RTPs in varying activity landscape satisfy TRS in position space. While computing the EPR in the full $(\br,\bu)$ space indeed leads to a positive number~\cite{razin2020entropy,cocconi2020entropy,frydel2022intuitive,paoluzzi2024entropy}, we show below that this apparent irreversibility disappears when only tracking the particle positions.

As for {OBP}s, we first show this at the level of trajectory. The algebra is much more involved and we do it only for RTP in $d=1$ before treating the general case at the operator level.

\subsubsection{TRS at the trajectory level for 1d RTPs}\label{sec:trs-abp-rtp-direct-calc}

Let us start by considering the dynamics of an RTP in one space dimension, whose dynamics reads
\begin{align}
    \dot{x} &= v(x) u(t) \,, \\
    u &\stackrel{{\tau^{-1}}}{\longrightarrow} u'\in \{-1, 1\}\;.
\end{align}
We note that the particle position evolves deterministically between tumbles and use this below to prove time-reversal symmetry in position space.

We consider a trajectory $\{ x(t) , u(t) \}$ in a time interval $t \in [t_\ri, t_\rf]$, with the boundary conditions $x(t_\mathrm{i,f}) = x_\mathrm{i,f}$ and $u(t_\mathrm{i,f}) = u_\mathrm{i,f}$. Without loss of generality, we simplify our notations by setting $t_\ri = 0$ and $t_\rf = \tau$. If the time evolution of the orientation $\{u(t)\}$ is specified, the probability density for the position evolves according to 
\begin{align} \label{eq:rtp_fpe}
    {\dot P}(x,t|\{u(t)\}, x_\ri) = -\partial_x [v(x) u(t) P] \;.
\end{align}
The solution to Eq.~\eqref{eq:rtp_fpe} is
\begin{align} \label{eq:rtp_fpe_solution}
    P (x,t|\{ u(t) \}, x_\ri ) = \delta[ x - X(t|x_\ri)] \;,
\end{align}
where $X(t|x_\ri)$ is the solution to $\dot{X}(t) = v(X) u(t)$ with $X(0) = x_\ri$, which can be verified directly by plugging Eq.~\eqref{eq:rtp_fpe_solution} into Eq.~\eqref{eq:rtp_fpe}.

Next, we consider the time-reversed evolution of the particle in position space, which is realized by reversing its orientation:
\begin{align}
    x(t) &\mapsto x^R(t)  = x(\tau - t) ~, \\
    u(t) &\mapsto u^R(t) = - u(\tau - t) ~.
\end{align}
Similar to Eq.~\eqref{eq:rtp_fpe_solution}, the probability density conditioned on the reversed orientation satisfies
\begin{align} \label{eq:rtp_fpe_tr_solution}
    P (x, t |\{ u^R (t) \}, x_\rf ) = \delta[x - X^R(t|x_\rf)]
\end{align}
where $X^R(t|x_\rf)$ is the solution of $\dot{X}^R(t) = v(X^R)u^R(t)$ with the initial position $X^R(0) = x_\rf$. 

Equation~\eqref{eq:rtp_fpe_solution} implies that
\begin{equation}
    P(x_\rf,\tau|\{ u(t) \},x_\ri)=\delta[g_\rf(x_\rf,x_\ri)]\;,
\end{equation}
with $g_\rf(x_\rf,x_\ri)=x_\rf-X(\tau|x_\ri)$.
Similarly, introducing $g_\ri(x_\rf,x_\ri)=x_\ri-X^R(\tau|x_\rf)$, Eq.~\eqref{eq:rtp_fpe_tr_solution} implies that
\begin{equation}
    P(x_\ri,\tau|\{ -u(t) \},x_\rf)=\delta[g_\ri(x_\rf,x_\ri)]\;.
\end{equation}
We note that $g_\rf$ and $g_\ri$ vanish on the same one-dimensional subset of $\mathbb{R}^2$ that we denote by $\mathcal{C}$. Let us now show that
\begin{equation}\label{eq:rtp1d:propagator}
   \frac{P (x_\ri, \tau| \{ u^R(t)\}, x_\rf )}{P (x_\rf , \tau | \{ u(t) \}, x_\ri) } = \frac{ v(x_\rf) }{v(x_\ri) } \;.
\end{equation}
Consider a function $f({\bf x}) \equiv f(x_\ri,x_\rf)$, then
\begin{align}\label{eq:toto}
    \int_{ \mathbb{R}^2 } f({\bf x})\delta[g_\rf({\bf x})] d{\bf x}&=\int_{\mathcal{C}} \frac{f({\bf x}) d\sigma({\bf x})}{|\nabla g_\rf|}
\end{align}
where $d\sigma({\bf x})$ is the Minkowsky content measure of $\mathcal{C}$ and $\nabla=(\partial_{x_\rf},\partial_{x_\ri})$. In turn, Eq.~\eqref{eq:toto} can be written as
\begin{align}
 \int_{\mathcal{C}} \frac{|\nabla g_\ri|}{|\nabla g_\rf|}\frac{f({\bf x})d\sigma({\bf x})}{|\nabla g_\ri|}
 =\int_{ \mathbb{R}^2 } f({\bf x})\frac{|\nabla g_\ri|}{|\nabla g_\rf|} \delta[g_\ri({\bf x})] d{\bf x}\label{eq:tutu}
\end{align}
Identifying Eqs.~\eqref{eq:toto} and~\eqref{eq:tutu} then leads to $\delta[g_\rf({\bf x})]=\delta[g_\ri({\bf x})]|\nabla g_\ri| /|\nabla g_\rf| $. Computing the gradients explicitly, we find
\begin{align}\label{eq:beautiful}
 \frac{P (x_\ri, \tau | \{ u^R(t)\}, x_\rf )}{P (x_\rf ,\tau | \{ u(t) \}, x_\ri) } =  \frac{ \sqrt{ 1 + (d X(\tau|x_\ri)/d x_\ri)^2} }{ \sqrt{ 1 + (d X^R(\tau|x_\rf)/d x_\rf)^2 } }.
\end{align}
\if{
Therefore, the ratio between \eqref{eq:rtp_fpe_solution} and \eqref{eq:rtp_fpe_tr_solution}, when $t = \tau$, leads to
\begin{align} \label{eq:rtp_P_ratio}
    \frac{P (x_\ri, \tau | \{ u^R(t)\}, x_\rf )}{P (x_\rf ,\tau | \{ u(t) \}, x_\ri) } = \frac{\delta[x_\ri - X^R(\tau|x_\rf)]}{\delta[ x_\rf - X(\tau|x_\ri)]}.
\end{align}
Due to the deterministic nature of the conditioned dynamics, the delta functions on the right-hand side are functions only of $x_\ri$ and $x_\rf$. Then we use the following relation that holds for delta functions:
\begin{align}
    \int_{ \mathbb{R}^n } f({\bf x}) \delta ( g({\bf x})) d{\bf x} = \int_{g({\bf x}) = 0} \frac{f({\bf x})}{|\nabla g|} d \sigma ({\bf x}),
\end{align}
where ${\bf x} \in \mathbb{R}^n$, $g({\bf x}) \in \mathbb{R}^n$, and the integration with $d\sigma({\bf x})$ on the right-hand side indicates an integration over the subspace of $\mathbb{R}^n$ where $g({\bf x}) = 0$. Rewriting Eq.~\eqref{eq:rtp_P_ratio} using the relation above while setting $\nabla = \hat{x}_\ri \partial_{x_\ri} + \hat{x}_\rf \partial_{x_\rf}$, we obtain
\begin{align}
    \nonumber \frac{P (x_\ri, \tau | \{ u^R(t)\}, x_\rf )}{P (x_\rf ,\tau | \{ u(t) \}, x_\ri) } = &= \frac{| \nabla [ x_\rf - X(\tau|x_\ri)] |}{|\nabla [x_\ri - X^R(\tau|x_\rf)] |} \\
    \label{eq:rtp_delta_refined} &= \frac{ \sqrt{ 1 + (d X(\tau|x_\ri)/d x_\ri)^2} }{ \sqrt{ 1 + (d X^R(\tau|x_\rf)/d x_\rf)^2 } }.
\end{align}
}\fi

To proceed, we rewrite $dX(t) / dt = v(X) u(t)$ as $dt = dX(t) / [v(X) u(t)]$, which we integrate to get
\begin{align} \label{eq:t_rtp}
    \tau = \int_{x_\ri}^{X(\tau)}  \frac{d x}{v(x) u[t(x)]}.
\end{align}
Differentiating Eq.~\eqref{eq:t_rtp} with respect to $x_\ri$ and rearranging leads to
\begin{align} \label{eq:dX/dxi}
    \left| \frac{dX(\tau|x_\ri)}{dx_\ri} \right| = \frac{v(x_\rf)}{v(x_\ri)}.
\end{align}
Similarly, for the time-reversed realization, we obtain 
\begin{align} \label{eq:dX/dxf}
    \left| \frac{dX^R(\tau|x_\rf)}{dx_\rf} \right| = \frac{v(x_\ri)}{v(x_\rf)}.
\end{align}
Plugging Eqs.~\eqref{eq:dX/dxi} and \eqref{eq:dX/dxf} into Eq.~\eqref{eq:beautiful}, we obtain Eq.~\eqref{eq:rtp1d:propagator}.

Since $\{u(t)\}$ follows a Poisson process with rate {$\tau^{-1}$} and the tumbles randomly sample new orientations isotropically, the probability density for the time-reversed realization satisfies $\mathbb{P}[\{u^R(t)\}] = \mathbb{P}[\{u(t)\}]$. The propagator from $x_i,u_i$ to $x_\rf,u_\rf$ is then found by summing over the realizations of the orientation trajectory:
\begin{align}
    \nonumber P&(x_\rf, u_\rf, \tau|x_\ri, u_\ri, 0) \\
    \nonumber &= \int {\cal D}[u(t)] \, \mathbb{P}[\{u(t)\}] P(x_\rf, \tau|\{u(t)\},x_\ri)\\
    &= \int {\cal D}[u(t)] \, \mathbb{P}[\{u^R(t)\}]  P(x_\ri,\tau|\{u^R(t)\},x_\rf) \frac{v(x_\ri)}{v(x_\rf)}\;,\notag
\end{align} 
where we have used Eq.~\eqref{eq:beautiful} in the last line and the path-integral over $u$ is constrained to $u(0)=u_\ri$ and $u(\tau)=u_\rf$. Since $u^R(t)\to u(t)$ is an involution, there is no Jacobian in the corresponding change of variables, and we find
\begin{equation} \label{eq:rtp_tr_final}
    P(x_\rf, u_\rf, \tau|x_\ri, u_\ri, 0) = P(x_\ri, -u_\ri, \tau| x_\rf, -u_\rf, 0) \frac{v(x_\ri) }{v(x_\rf)}.
\end{equation}

Let us now show that this implies TRS in position space. Considering that the system is in the steady state at time $0$, the probability to be in $x_\ri$ at time 0 and in $x_\rf$ at time $\tau$ is given by
\begin{align}
    P(x_\rf,\tau;x_\ri,0)&=\sum_{u_\rf,u_\ri} P(x_\rf,u_\rf,\tau;x_\ri,u_\ri,0)\notag\\
    &=\sum_{u_\rf,u_\ri} P(x_\rf,u_\rf,\tau|x_\ri,u_\ri,0) \psi_{\rm s}(x_\ri,u_\ri)\;,
\end{align}
where we have used the definition of a conditional probability and $\psi_{\rm s}$ is the steady state measure. Using Eq.~\eqref{eq:rtp_tr_final}, one then find
\begin{align}
    P(x_\rf,\tau;x_\ri,0)=\sum_{u_\rf,u_\ri} P(x_\ri,-u_\ri,\tau|x_\rf,-u_\rf,0)  \psi_{\rm s}(x_\ri,u_\ri)\frac{v(x_\ri) }{v(x_\rf)}\;.\nonumber
\end{align}
Using that $\psi_{\rm s}(x,u)\propto 1/v(x)$, we then find
\begin{align}
    P(x_\rf,\tau;x_\ri,0)&=\sum_{u_\rf,u_\ri} P(x_\ri,-u_\ri,\tau|x_\rf,-u_\rf,0)  \psi_{\rm s}(x_\rf,-u_\rf)\notag\\
    &=P(x_\ri,\tau;x_\rf,0)\;.
\end{align}
The dynamics is thus time-reversal symmetric in position space.

\subsubsection{RTPs and ABPs in $d$ dimensions}\label{sec:trs-abp-rtp-operator}
Let us now show the existence of TRS in position space using an operator approach for RTPs and ABPs in $d$ dimensions. Consider Eqs.~\eqref{eq:ri_abp_rtp} and~\eqref{eq:fpe-abp-rtp} in the absence of interactions. The corresponding Master equation can be written as
\begin{align}
    \dot{\psi}(\br,\bu,t) &= -\nabla \cdot \left[v(\br) \bu \psi\right] + \mathcal{L}_\mathrm{r} \psi \equiv {\cal L} \psi \label{eq:abp-rtp-fpe-ni}
\end{align}
where we have introduced the evolution operator \begin{equation}
    {\cal L}=-\nabla \cdot v(\br) \bu + \mathcal{L}_\mathrm{r}(\bu)\;,
\end{equation}
and $\mathcal{L}_\mathrm{r}$ is the Hermitian evolution operator of the angular dynamics. The adjoint evolution operator reads ${\cal L}^\dag = v(\br)\bu\cdot \nabla + \mathcal{L}_\mathrm{r}(\bu)$. We remind that, in the non-interacting case, the steady-state solution satisfies $\psi_\mathrm{s}(\br,\bu) = \psi_\mathrm{s}(\br,-\bu)=\kappa/v(\br)$, where $\kappa$ is determined by normalization.

Direct algebra shows that
\begin{equation}\label{eq:rtp-nd-TRS-op}
    \mathcal{L}^\dagger=  \psi_{\rm s}^{-1} \Pi^{-1} \mathcal{L} \Pi  \psi_{\rm s}\;,
\end{equation}
where the operator $\Pi$ is such that with $\Pi f(\br,\bu)=f(\br,-\bu)$. Let us now show that Eq.~\eqref{eq:rtp-nd-TRS-op} implies TRS in position space. We consider that the system is in the steady state at time 0 and compute $   P(\br',t,\br,0)$, the probability to be at $\br$ at time $0$ and $\br'$ at time $t$. Using Dirac bra-ket notations for the propagator, we find
\begin{align}
   P(\br',t,\br,0)&=\int d\bu d\bu' P(\br',\bu,'t,\br,\bu,0)\notag\\
    &=\int d\bu d\bu' \langle \br',\bu' | e ^{t \mathcal{L}}  | \br,\bu \rangle  \psi_{\rm s}(\br,\bu)  \notag\\
    &=\int d\bu d\bu'  \langle \br,\bu | e ^{t \mathcal{L}^\dagger} | \br',\bu' \rangle \psi_{\rm s}(\br,\bu)\;,\label{eq:rtp-nd-propa1}
\end{align}
where, in the last equality, we have used that the propagator $ \langle \br',\bu' | e ^{t \mathcal{L}}  | \br,\bu \rangle$ is a real number and hence equal to its adjoint. Equation~\eqref{eq:rtp-nd-TRS-op} then tells us that
\begin{align}
   \langle \br,\bu | e ^{t \mathcal{L}^\dagger } | \br',\bu' \rangle&=\langle \br,\bu |\psi_{\rm s}^{-1} \Pi^{-1} e ^{t \mathcal{L}}\Pi \psi_{\rm s} | \br',\bu' \rangle\notag\\
   &=\frac{\psi_{\rm s}(\br',\bu')}{\psi^*_{\rm s}(\br,\bu)}\langle \br,\bu | \Pi^{-1} e ^{t \mathcal{L}}\Pi | \br',\bu' \rangle\notag\\
   &=\frac{\psi_{\rm s}(\br',\bu')}{\psi_{\rm s}(\br,\bu)}\langle \br,-\bu | e ^{t \mathcal{L}} | \br',-\bu' \rangle\;.\label{eq:rtp-nd-propa2}
\end{align}
Equations~\eqref{eq:rtp-nd-propa1}  and~\eqref{eq:rtp-nd-propa2} then lead to
\begin{align}
    P(\br',t,\br,0)&=\int d\bu d\bu' \psi_{\rm s}(\br',\bu') P(\br,-\bu,t | \br',-\bu',0)\notag\\
    &=P(\br,t,\br',0)\;,
\end{align}
where we have used that $\psi_{\rm s}(\br',\bu')=\psi_{\rm s}(\br',-\bu')$ in the final line.

Thus, ABPs and RTPs with spatially varying self-propulsion obey TRS in position space and cannot experience {directed} currents.

\subsubsection{Interaction-induced TRS violations}
Interacting active matter has frequently been shown to exhibit nonzero entropy production in position space~\cite{fodor_how_2016,martin_statistical_2021,ro_model_2022,obyrne_time_2022,mallmin2019exact}. In fact, for ABPs and RTPs without translational diffusivity, pairwise forces typically make the dynamics fully irreversible{: the time-reversal of a trajectory is typically not a solution of the equations of motion, whatever the realizations of the noises that determine the self-propulsion forces. 

To show this, consider} the interacting dynamics Eq.~\eqref{eq:ri_abp_rtp}:
\begin{align}
    \dot{\br}_i(t) &= \mathbf{F}_i(\{\br_j(t)\}) + v(\br_i(t)) \bu_i(t)\;.
    \label{eq:active_trs_forward}
\end{align}

We note that, for the time-reversed realization of $\br$ to be solution of the dynamics, it needs to satisfy
\begin{equation}
    \nonumber \dot{\br}_i^R(t')= \mathbf{F}_i(\{\br_j^R(t')\}) + v(\br_i^R(t')) \bu_i^R(t')\nonumber\;,
\end{equation}
{with $\bu_i^R(t')$ a unit vector. }
Then, using that ${\br}_i(t)={\br}_i^R(\tf-t)$ and $\dot{\br}_i(t)=-\dot{\br}_i^R(\tf-t)$, this implies
\begin{align}
-\dot{\br}_i(t) &= \mathbf{F}_i(\{\br_j(t)\}) + v(\br_i(t)) \bu_i^R(\tf-t)\;.
    \label{eq:active_trs_reverse}
\end{align}
Adding Eqs.~\eqref{eq:active_trs_forward} and \eqref{eq:active_trs_reverse} then leads to
\begin{align}
    0 &= 2 \mathbf{F}_i(\{\br_j(t)\}) + v(\br_i(t)) \left[ \bu_i^R(\tf-t) + \bu_i(t)\right].
    \label{eq:active_trs_condition}
\end{align}
If $\mathbf{F}_i(\{\br_j(t)\}) = 0$, Eq.~\eqref{eq:active_trs_condition} can be satisfied by setting ${\bf u}_i^R(t_\mathrm{f} - t) = - {\bf u}_i(t)$; for a noninteracting system a time-reversed trajectory can be realized by flipping the orientation {of the forward trajectory}. 

On the contrary, when $\mathbf{F}_i(\{\br_j(t)\}) \neq 0$, it is impossible to satisfy Eq.~\eqref{eq:active_trs_condition} generically because {it requires}
\begin{align}\label{Eq:fullirreversibility}
    {\big|\bu_i^R(\tf-t)| = \bigg|-\frac{2\mathbf{F}_i(\{\br_j(t)\})}{v(\br_i(t))} - \bu_i(t)\bigg|\;.}
\end{align}
{For ${\br}_i^R(\tf-t)$ to solve the equations of motion, we need to have $|\bu_i^R(\tf-t)|=1$. But the right-hand side of Eq.~\eqref{Eq:fullirreversibility} is generically not a unit vector. The time-reversal of the trajectories of interacting ABPs or RTPs thus generically do not correspond to trajectories of interacting ABPs or RTPs, regardless of the choice of self-propulsion noises. Such time-reversed trajectories are thus not merely less likely than the forward trajectories, they simply never occur.}

\subsection{Momentum sources}\label{sec:abp-rtp-eos}
Let us now show that, for interacting ABPs and RTPs in the presence of an activity landscape, momentum sources arise that allow the emergence of {directed} currents. 

To do so, we start from Eq.~\eqref{eq:fpe-abp-rtp} for the empirical distribution. We use the standard method of decomposing $\psi(\br,\bu)$ into a hierarchy of orientational modes given by the density $\hat{\rho}(\br) = \sum_i \delta(\br-\br_i)$, magnetization vector $\hat{\bfm}(\br) = \sum_i \bu_i \delta(\br-\br_i)$, nematic order tensor $\hat{\bQ}(\br) = \sum_i \delta(\br-\br_i) \left(\bu_i \otimes \bu_i - \frac{\mathbb{I}}{d}\right)$, and so on~\cite{cates_when_2013}. For both ABPs and RTPs, we obtain for the average fields~\cite{solon_pressure_2015-1,solon2018generalized,solon2018generalizedNJP,speck2021coexistence,wysocki_interacting_2022,omar2023mechanical}
\begin{align}
    \dot \rho &= -\nabla \cdot \bigg[ \bfm v - \int\hspace{-0.25em} d\br' \nabla \U(\br-\br') \langle\hat{\rho}(\br)\hat{\rho}(\br')\rangle\bigg] \, ,\label{eq:rhodot-abp-rtp}\\
    \dot \bfm &= -\nabla \cdot \bigg[v \left(\bQ + \mathbb{I} \frac{\rho}{d}\right)  \nonumber\\
    &\quad- \int d\br' \nabla \U(\br-\br') \otimes \langle \hat{\bfm}(\br)\hat{\rho}(\br')\rangle\bigg] - \frac{\bfm}{\tau} \, .\label{eq:mdot-abp-rtp}
\end{align}
(The dynamics of higher moments is not required here.) In steady state,
we set ${\dot\bfm}= 0$ to get
\begin{equation}
 \bfm    \equiv -\tau \nabla \cdot \bJ_m\;,
\end{equation}
where
\begin{align}
    \bJ_m&= v \left(\bQ + \mathbb{I} \frac{\rho}{d}\right)  - \int d\br' \nabla \U(\br-\br') \otimes \langle \hat{\bfm}(\br)\hat{\rho}(\br')\rangle
    \end{align}
Making this substitution, and absorbing the mechanical forces in Eq.~\eqref{eq:rhodot-abp-rtp} into the Irving-Kirkwood stress $\bsigma_\subIK$~\cite{irving1950statistical}, we find
\begin{align}
    \dot \rho &= -\nabla \cdot \big\{-\tau v(\br) \nabla \cdot \bJ_m + \nabla \cdot \bsigma_\subIK \big\}\nonumber\\
    &= -\nabla \cdot \big\{\nabla \cdot \big[-\tau v \bJ_m + \bsigma_\subIK\big] + \tau [\nabla v(\br)] \cdot \bJ_m\big\}\nonumber\\
    &\equiv -\nabla \cdot \big\{\nabla \cdot \bsigma_{\rm tot} + \delta \bF_\subA\big\}\;.
    \label{eq:abp-rtp-no-eos}
\end{align}
If the activity is constant, the active forces can be written as the divergence of an active stress, as expected~\cite{solon_pressure_2015-1}. However, variations in the self-propulsion speed lead to a nonconservative active force $\delta \bF_A{=\tau [\nabla v(\br)] \cdot \bJ_m}$, as identified in~\cite{wysocki_interacting_2022}. These forces cannot be written as the divergence of a local stress tensor, thus generating sinks and sources of momentum which, as we show below, can power steady-state {directed} currents.

\subsection{Interaction-induced current}
Interacting ABPs and RTPs in asymmetric activity landscapes thus violate TRS, parity symmetry, and momentum conservation, which generically allows for {directed} currents. Let us now show how we can predict the emergence of such currents quantitatively using a perturbation theory in the interaction potential to solve Eqs.~\eqref{eq:rhodot-abp-rtp}-\eqref{eq:mdot-abp-rtp}. 

First, to make progress we use the factorization approximation
\begin{align}
    \langle \hat{\rho}(\br) \hat{\rho}(\br')\rangle &= \langle \hat{\rho}(\br)\rangle \langle \hat{\rho}(\br')\rangle =  {\rho}(\br)  {\rho}(\br')\label{eq:mf-rho-abp-rtp}\\
    \langle \hat{\bfm}(\br) \hat{\rho}(\br')\rangle &= \langle \hat{\bfm}(\br)\rangle \langle \hat{\rho}(\br')\rangle = \bfm(\br) \rho(\br')\;.\label{eq:mf-m-abp-rtp}
\end{align}
In this Curie-Weiss mean-field picture, the particles at position $\br$ experience an effective potential $V_\eff(\br) \equiv \U\ast\rho(\br)$ as in Eq.~\eqref{eq:Veff-convolution-od-pbp}. Then, the dynamics boil down to those for an ABP/RTP in $d$ dimensions moving in the activity landscape $v(\br)$ and the potential landscape $V_{\rm eff}(\br)$. However, since the latter depends on $\rho$, these equations are nonlinear and should be solved self-consistently. As we now show, progress can be made in the limit of a weak interaction potential. We thus rewrite $\U=\varepsilon\tU$ and work perturbatively in $\varepsilon$.

\subsubsection{Leading-order {directed} current for ABPs and RTPs in $d$ dimensions}\label{sec:abp-rtp-perturbative-J}
The mean-field steady-state equations for ABPs and RTPs in $d$ dimensions read $\nabla\cdot\bJ=0$ with
\begin{align}
\bJ &= \bfm v - \varepsilon\rho \nabla (\tU \ast \rho),\quad  \label{eq:mf-rhodot-abp-rtp}\\
    \bfm &= - \tau \nabla \cdot \Big\{v \Big(\bQ + \mathbb{I} \frac{\rho}{d}\Big) - \varepsilon\bfm \otimes \nabla (\tU\ast \rho) \Big\}\;.\label{eq:mf-mdot-abp-rtp}
\end{align}
 We then consider a perturbative ansatz for all fields
\begin{align}
    \rho &= \sum_{k=0}^\infty \varepsilon^k \rho_k\;,\quad \bfm = \sum_{k=0}^\infty \varepsilon^k \bfm_k\;,\quad \bQ = \sum_{k=0}^\infty \varepsilon^k \bQ_k\;,\,\ldots\nonumber\\
    \bJ &= \sum_{k=0}^\infty \varepsilon^k \bJ_k
\end{align}

At zeroth order in $\varepsilon$, we find $\bJ_0=\bfm_0=0$ and {$\rho_0=\kappa/v$ with $\kappa$ a normalization constant}, consistent with the result for noninteracting particles. At first order, Eqs.~\eqref{eq:mf-rhodot-abp-rtp}-\eqref{eq:mf-mdot-abp-rtp} imply
\begin{align}
    \bJ_1 &= \bfm_1 v - \rho_0 \nabla (\tU \ast \rho_0)\;,\label{eq:mf1-rhodot-abp-rtp}\\
    \bfm_1 &= - \tau \nabla \cdot \Big\{v \Big(\bQ_1 + \mathbb{I} \frac{\rho_1}{d}\Big)\Big\}\;.\label{eq:mf1-mdot-abp-rtp}
\end{align}
These equations are not closed, since they involve $\bQ_1$. To make progress, we consider an effectively 1-dimensional activity landscape $v(\br)=v(x)$, as in Fig.~\ref{fig:abp-rtp-2d-phenomenology}. Then, by symmetry, the steady-state current has to satisfy $\bJ_1=J_1 \hat x$, with $J_1$ a constant. Dividing Eq.~\eqref{eq:mf1-rhodot-abp-rtp} by $v(x)$ and integrating over space then leads to
\begin{align}
    J_1 &= -\kappa^3 \int_0^L dx \int_0^L dx' \frac{\Tilde{U}_{\rm int}'(x-x')}{v(x)^2 v(x')}\label{eq:abp-rtp-J1}\;.
\end{align}
\if{\begin{align}
    \rho_1 &= \frac{\omega_1}{v} - \frac{d}{\tau v} \int_0^x\hspace{-0.5em} \frac{dx'}{v(x')} \left[J_1 + \frac{\omega_0^2 (\U\ast v^{-1})'(x')}{v(x')}\right]\label{eq:abp-rtp-rho1}\\
    J_1 &= -\omega_0^3 \int_0^L\hspace{-0.5em} dx \int_0^L \hspace{-0.5em} dx' \frac{\U'(x-x')}{v(x)^2 v(x')}
\end{align}}\fi
This prediction is plotted as the black curve in Fig.~\ref{fig:abp-rtp-2d-phenomenology}(c) and compared to the result of numerical simulations. While it predicts the order of magnitude of $J$ when the interaction strength is weak, the slope at $\varepsilon=0$ slightly differs from that measured numerically, probably due to the factorization approximation in~\eqref{eq:mf-rho-abp-rtp}-\eqref{eq:mf-m-abp-rtp}. It also misses the current reversal at larger values of $\varepsilon$.

Going to higher order in $\varepsilon$ requires a closure in the moments of $\psi$, which can be done by taking a small $|\nabla v|$ limit. Alternatively, we here consider one-dimensional run-and-tumble particles for which the expansion can be carried out to arbitrary order without further approximations.

\subsubsection{Run-and-Tumble Particles in 1D}\label{sec:1d-rtps}
We here detail the perturbation theory for the case of  RTPs in $d=1$, whose dynamics read
\begin{align}\label{eq:xdot-1d-rtp}
    \dot{x}_i &= v(x_i) u_i - \sum_j \U'(x_i-x_j)\;,
\end{align}
where the orientation $u_i \in \{-1,+1\}$ is randomized at rate $\tau^{-1}$. These dynamics can be equivalently written for density and magnetization fields $\hat{\rho}(x) = \sum_i \delta(x-x_i) $ and $\hat{m}(x) = \sum_i u_i \delta(x-x_i)$ as coupled hydrodynamic equations
\begin{align}
    \dot\rho &= -\partial_x \Big[v m - \int dx' \langle \hat{\rho}(x) \hat{\rho}(x')\rangle \U'(x-x')\Big] \, ,\label{eq:dynr1}\\
    \dot m &= -\partial_x \Big[v \rho - \int dx' \langle \hat{m}(x) \hat{\rho}(x')\rangle \U'(x-x')\Big]- \frac{m}{\tau}\;.
\end{align}
Performing the mean-field factorization Eqs.~\eqref{eq:mf-rho-abp-rtp}-\eqref{eq:mf-m-abp-rtp} then yields
\begin{align}
    \dot\rho &= -\partial_x \Big[v m - \rho \partial_x (\U \ast \rho)\Big]\label{eq:rtp-1d-int-rhodot}\\
    \dot m &= -\partial_x \Big[v \rho - m \partial_x (\U\ast\rho)\Big]- \frac{m}{\tau}.
    \label{eq:rtp-1d-int-mdot}
\end{align}
The equations above are equivalent to RTP equations within an effective potential $V_\eff \equiv \U\ast\rho$. In fact, the steady-state for 1d RTPs in an arbitrary \textit{external} force $f(x)$ (e.g., $f(x)=-V'(x)$) can be solved explicitly in the absence of interactions as detailed in Appendix~\ref{appendix:rtp-1d-van-kampen}. As for {OBP}s, the resulting solution could then be used to determine the steady-state for \textit{interacting} system by inserting $f(x)=-\partial_x V_\eff = -\partial_x (\U\ast\rho)$, and performing a self-consistent perturbation in the strength of $\U$. 

Here, we follow the more direct route of solving~Eqs.~\eqref{eq:rtp-1d-int-rhodot} and~\eqref{eq:rtp-1d-int-mdot} perturbatively. We again denote the interaction potential as $\U=\varepsilon \tU$ and expand the fields in powers of the $\varepsilon$:
\begin{align}
    \rho = \sum_{k=0}^\infty \varepsilon^k \rho_k,\quad m = \sum_{k=0}^\infty \varepsilon^k m_k,\quad J = \sum_{k=0}^\infty \varepsilon^k J_k\;.
\end{align}
We then solve the steady-state equations
\begin{align}
    J &= v m - \varepsilon \rho \partial_x (\tU\ast\rho)\;,\quad J \;\text{constant} \; ,\\
    m &= -\tau \partial_x \Big[v \rho - \varepsilon m \partial_x (\tU \ast\rho)\Big]
\end{align}
order-by-order in $\varepsilon$.

At 0th order, the solution reads
\begin{align}
    \rho_0(x) = \frac{\kappa_0}{v(x)},\quad m_0(x) = 0,\quad J_0=0
\end{align}
where $\kappa_0 = N \big[\int_0^L  \frac{dx}{v(x)}\big]^{-1}$ ensures $\rho_0$'s normalization.

At 1st order, we find
\begin{align}
    \rho_1 &= \frac{1}{\tau v} \left\{\kappa_1 - \int_0^x \frac{dx}{v(x')} \left[J_1 + \frac{\kappa_0^2(\tU\ast v^{-1})'(x')}{v(x')}\right]\right\} \,,\notag\\
    m_1 &= -\tau \partial_x (v \rho_1)\,,\notag\\
    J_1 &= -\kappa_0^3 \int_0^L dx \int_0^L dx' \frac{\tU'(x-x')}{v(x)^2 v(x')}\;,\notag
\end{align}
in agreement with Eq.~\eqref{eq:abp-rtp-J1}. Once again, a {directed} current appears for a generic asymmetric $v(x)$. 

\begin{figure}
    \centering
    \begin{tikzpicture}
    \path (-0,0) node {\includegraphics{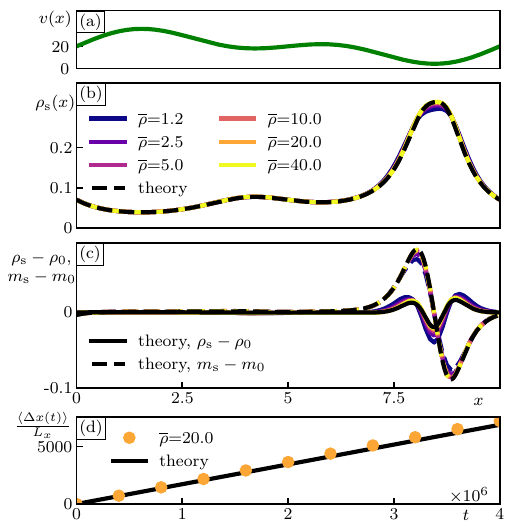}};
    \end{tikzpicture}
    \caption{\textbf{Interacting RTPs in 1d.} {\bf (a)}  Activity landscape $v(x)$. \textbf{(b)} Steady-state density profile $\rho_{\rm s}(x)$ as the average density $\bar\rho$ is varied. The dashed black line corresponds to the theoretical prediction to order {3} in the perturbation theory. \textbf{(c)} Detailed comparison of the convergence between numerics and theory as $\bar \rho$ is increased for the density (solid) and orientation (dashed) fields. \textbf{(d)} Average displacement per particle (symbol) compared to the theoretical prediction (solid line). Simulation details are given in Appendix~\ref{appendix:sim-abp-rtp}.  }
    \label{fig:1d-rtp-interact-profile-disp}
\end{figure}

This calculation can be carried to arbitrary order using the recursive relation
\begin{align}
    J_n &= -\kappa_0 \sum_{k=0}^{n-1} \int_0^L dx \frac{\rho_k(x) (\tU\ast\rho_{n-k-1})'(x)}{v(x)}\,,\notag\\
    {m_n }&{= \frac{J_n}{v} + \frac{1}{v} \sum_{k=0}^{n-1} \rho_k (\tU\ast\rho_{n-k-1})'}\,,\notag\\
    {\rho_n }&{=  \frac{1}{v}\bigg[ \kappa_n + \sum_{k=0}^{n-1} m_k (\tU\ast\rho_{n-k-1})' - \int_0^x dx'  \frac{m_n}{\tau}\bigg]}\;.\notag
\end{align}
The result of this procedure is illustrated in Fig.~\ref{fig:1d-rtp-interact-profile-disp} {(to order 3)}. It agrees very well with numerical simulations in the high-density limit where the factorization approximation is expected to work best.

\subsection{Summary}

ABPs and RTPs in parity-breaking activity landscapes do not exhibit steady-state currents in the absence of interactions. While this appears to violate the ratchet principle, since active matter is out of thermal equilibrium, we have shown that the particles obey detailed balance with respect to their non-Boltzmann steady-state measures in position space. This effective TRS {in position space} thus forbids a {directed} current in the non-interacting case.

In the presence of interactions, TRS is violated and the systems do not admit a generalized stress tensor, so that nonconservative forces can act as momentum sources, leading to steady-state currents. We have seen these currents borne out numerically, and characterized their emergence analytically employing a mean-field factorization approximation (Figs.~\ref{fig:abp-rtp-2d-phenomenology} and~\ref{fig:1d-rtp-interact-profile-disp}).

In the mean-field picture, the mechanism behind the interaction-induced current is the coupling between the activity landscape and the effective potential that each particle experiences from its  neighbors.

\section{Active Ornstein-Uhlenbeck Particles}\label{sec:aoups}

\begin{figure*}
    \begin{tikzpicture}
    \path (-0,0) node {\includegraphics{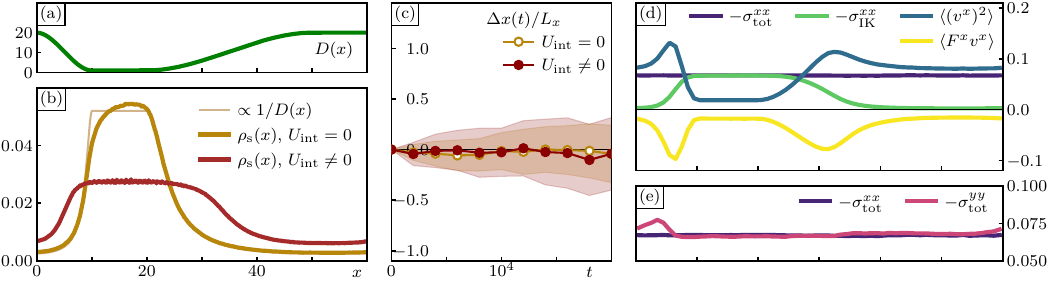}};
    \end{tikzpicture}
    \caption{\textbf{AOUPs {in an activity landscape} are protected from a {directed} current by an emergent momentum conservation.} {\bf (a)} The activity landscape $D(x)$ used in simulations. {\bf (b)} The steady-state density distribution of AOUPs both without interactions (gold) and with interactions (dark red). The density in the $\tau\rightarrow 0$ limit, $\propto 1/D(x)$, is shown for reference (thin beige line). {\bf (c)} Directly measuring the net particle displacement over time, both without and with interactions, reveals the absence of {directed} current. {\bf (d)} Remarkably, AOUPs possess steady-state momentum conservation, so that gradients of active stresses $\tau \langle \bv_i \otimes \bv_i\rangle$ (blue) and $\tau \langle \bF_i \otimes \bv_i\rangle$  (yellow) are balanced by gradients in the Irving-Kirkwood stress (green). {\bf (e)} The $(y,y)$ stress $\sigma_{\rm tot}^{yy}$ is spatially varying unlike $\sigma_{\rm tot}^{xx}$; however it is constant along $y$ by translational symmetry, thus generating no current. 
    Simulation details are provided in Appendix~\ref{appendix:sim-aoup}.
    }
    \label{fig:aoup-eos}
\end{figure*}

Finally, we consider another frequently-studied active model called Active Ornstein-Uhlenbeck Particles (AOUPs), which, as we show below, offers interesting differences with respect to ABPs and RTPs. We consider $N$ particles evolving in $d$ dimensions according to~\cite{martin_statistical_2021}
\begin{align}
    \dot{\br}_i &= \bv_i - \sum_j \nabla \U(\br_i-\br_j) \; ,\label{eq:langevin-aoup-a}\\
    \tau \dot{\bv}_i &= - \bv_i + \sqrt{2 D(\br
    _i) } \bfeta_i(t) \; ,
    \label{eq:langevin-aoup-b}
\end{align}
where $\bfeta_i(t)$ is a centered Gaussian white noise satisfying $\langle \eta_{i, \mu}(t)\rangle=0$ and $\langle \eta_{i,\mu}(t) \eta_{j,\nu}(t')\rangle = \delta_{ij} \delta_{\mu\nu} \delta(t-t')$. In Eq.~\eqref{eq:langevin-aoup-b}, the large-scale diffusivity $D({\bf r})$ makes the AOUP activity vary in space. 

We consider the empirical density of particles in the position-propulsion-velocity space $\hat{\psi}(\br,\bv,t) \equiv \sum_i \delta(\br_i-\br) \delta(\bv_i-\bv)$. The equation describing the dynamics of its average, $\psi(\br,\bv)=\langle \hat{\psi}(\br,\bv)\rangle$, is
\begin{align} 
    \dot{\psi}&(\br,\bv,t) = \frac{1}{\tau} \nabla_\bv \cdot \left[\bv \psi + \frac{D(\br)}{\tau} \nabla_\bv \psi\right] - \nabla \cdot \bigg[ \bv \psi \nonumber \\
    & - \int d\br' \int d\bv' \nabla \U(\br-\br') \langle \hat{\psi}(\br,\bv) \hat{\psi}(\br',\bv')\rangle \bigg]\;. \label{eq:kramers-aoup} 
\end{align}

{Directed} currents have been shown to be forbidden for noninteracting AOUPs in one-dimensional activity landscapes $D(x)$~\cite{martin_statistical_2021}. Noting that TRS is violated for AOUPs with spatially-varying activity, both with and without interactions~\footnote{This is a direct consequence of the mapping of non-interacting AOUPs onto the {UBP}s considered in Sec.~\ref{sec:ud-pbp-trs}.}, we demonstrate that, somewhat surprisingly, interaction-induced {directed} currents are forbidden in AOUPs. As for {UBP}s, this stems from an effective bulk momentum conservation. To demonstrate these results, we provide an exact expression for the generalized stress tensor, which forbids the momentum sources required to generate steady-state {directed} currents.

\subsection{Numerics and phenomenology}\label{sec:aoup-numerics-phenom}
The results of particle-based simulations of AOUPs in $d=2$ space dimensions, with an effectively one-dimensional activity landscape $D(x)$, are reported in Fig.~\ref{fig:aoup-eos}. The density distribution in the non-interacting case is distinct from its $\tau\to 0$ limit $\rho_{\rm s}(x) \propto 1/D(x)$. It is further deformed by adding repulsive interactions between the particles, which flatten the density~[Fig.~\ref{fig:aoup-eos}(b)]. There is no {directed} current in interacting AOUPs, as evidenced by tracking the integrated particle displacements over time~[Fig.~\ref{fig:aoup-eos}(c)]. The generalized stress tensor of AOUPs is measured and characterized in Fig.~\ref{fig:aoup-eos}(d-e) to illustrate the derivations of Sec.~\ref{sec:aoup-eos}.

\subsection{Time-reversal symmetry}\label{sec:aoup-trs}
{We note that the definition of entropy-production rate for AOUPs has been the topic of a longstanding debate~}\cite{fodor_how_2016,mandal_entropy_2017,caprini_comment_2018,mandal_mandal_2018,dabelow_irreversibility_2019}{. 
Here, we stick to our information-theoretic definition of $\sigma$, and stay clear of the discussions on its thermodynamic meaning, so that there is no ambiguity on the studied observable, nor on its relationship to currents and irreversibility in position space.}

In the absence of interactions, the dynamics for AOUPs [Eqs.~\eqref{eq:langevin-aoup-a}-\eqref{eq:langevin-aoup-b}] is identical to the dynamics for {UBP}s [Eq.~\eqref{eq:langevin-ud-pbp}], if $\tau$ is replaced by $1/\gamma$ and $D(\br)$ is replaced by $T(\br)/\gamma$. As discussed in Sec.~\ref{sec:ud-pbp-trs}, {UBP}s with spatially-varying temperature have nonzero EPR. The same thus holds for the equivalent system of AOUPs with spatially varying $D$. Consequently, AOUPs are {also} ``genuine" exceptions to the ratchet principle, violating both TRS in position space and parity symmetry while lacking a current.

Like all other systems considered in this paper, adding interactions creates new mechanisms for entropy production. We calculate this directly in Appendix~\ref{appendix:path-integral-aoup}, where we show the EPR field to be given by:
\begin{align}
    \hat{\sigma}(\br) &= \sum_{i} \frac{\delta(\br-\br_i)}{D(\br_i)} \Big[\sum_j \bfint(\br_i-\br_j) - \tau \ddot{\br}_i \Big] \label{eq:aoup-epr-density}\\
    &\quad\quad\quad \cdot \Big[\dot{\br}_i - \tau \sum_j (\dot{\br}_i-\dot{\br}_j)\cdot \nabla \bfint(\br_i-\br_j)\Big]\;.\nonumber
\end{align}
Note that the last term of the second bracket does not appear in the EPR field of {UBP}s, Eq.~\eqref{eq:ud-pbp-epr-density}, and is thus unique to interactions between AOUPs.

\subsection{Momentum conservation}\label{sec:aoup-eos}
We now project Eq.~\eqref{eq:kramers-aoup} onto a hierarchy of orientational modes given by the density $\hat{\rho}(\br) = \sum_i \delta(\br-\br_i)$, magnetization vector $\hat{\bfm}(\br) = \sum_i \bv_i \delta(\br-\br_i)$, nematic order tensor $\hat{\bQ}(\br) = \sum_i \delta(\br-\br_i) \left(\bv_i \otimes \bv_i - \mathbb{I}\frac{D(\br_i)}{\tau}\right)$. As a result, we find
\begin{align}
    \dot{\rho} &= -\nabla \cdot \bigg[\bfm\label{eq:aoup-rhodot} + \int d \br'\bfint(\br-\br') \langle\hat{\rho}(\br)\hat{\rho}(\br')\rangle\bigg]\\
    \dot{\bfm} &= -\nabla \cdot \mathbf{J}_m - \bfm/\tau,\quad
\end{align}
where 
\begin{align}
    \mathbf{J}_m &\equiv \mathbf{Q} + \mathbb{I} \frac{\rho D}{\tau} + \big\langle \hat{\bfm} \otimes \left(\bfint\ast \hat{\rho}\right)\big\rangle \;.
\end{align}
In the steady state, $\dot{\bf m} = 0$ so that ${\bf m} = -\tau \nabla \cdot {\bf J}_m$. By recognizing that Eq.~\eqref{eq:aoup-rhodot} has the structure of a continuity equation, $\dot{\rho} = -\nabla \cdot {\bf J}$,  we express the steady-state current as the divergence of a stress tensor:
\begin{align}
    \bJ &= -\nabla \cdot [ \tau {\bf J}_m - \bsigma_{\rm IK}] = \nabla \cdot \bsigma_{\rm tot}
    \label{eq:aoup-eos} 
\end{align}
where $\bsigma_{\rm IK}$ is the Irving-Kirkwood stress tensor~\cite{irving1950statistical}. {Thus, using the divergence theorem, $\int d \br \bJ=0$ and there is no {directed} current in the steady state.}

In Fig.~\ref{fig:aoup-eos}(d), we plot the $(x,x)$ component of the AOUPs stress tensor measured in particle-based simulations. From Eq.~\eqref{eq:aoup-eos}, we obtain
\begin{align}
    \nonumber \bsigma_{\rm tot}^{xx} = \bsigma_{\rm IK}^{xx} - \sum_i \left[ \tau \langle v_{i}^x \otimes v_{i}^x \rangle + \tau \langle v_{i}^x \otimes {F}_{i}^x \rangle \right]\;,
\end{align}
where $F_i^x=-\sum_j \partial_{x_i}\U(\br_i-\br_j)$ is the force experienced by particle $i$ along $\hat x$.
Our measurements show that the active stresses, $-\tau \langle \bv_i \otimes \bv_i\rangle$ and  $-\tau \langle \bF_i \otimes \bv_i\rangle$, and the passive one, $\bsigma_{\rm IK}$, balance each other, resulting in a constant total stress and a flux-free steady state.
Interestingly, the $(y,y)$ component of the total stress shown in Fig.~\ref{fig:aoup-eos}(e) is not constant with respect to $x$. This does not affect the current since $\bJ = \hat{x} \partial_x \bsigma_{\rm tot}^{xx} + \hat{y} \partial_y \bsigma_{\rm tot}^{yy}$ and $\partial_y \bsigma_{\rm tot}^{yy}=0$.

Our generalized stress tensor is also useful to pinpoint similarities and differences between AOUPs and {UBP}s. Comparing Eq.~\eqref{eq:aoup-eos} with the {UBP} stress tensor~\eqref{eq:ud-pbp-J-eos}, we find
\begin{align}
    \bsigma_{{\rm \text{{UBP}}}} &= \frac{1}{\gamma} \left[\bsigma_\subIK - \bQ - \mathbb{I} \rho T \right]\;,\\
    \bsigma_\subAOUP &= \tau \left[\bsigma_\subIK - \bQ - \mathbb{I} \frac{\rho D}{\tau} - \big\langle \hat{\bfm} \otimes (\bfint \ast \rho)\big\rangle \right]\;.
    \label{eq:ud-pbp-aoup-stress-comparison}
\end{align}
They are almost identical under the mapping that identifies $\bp \leftrightarrow \bv$,  $1/\gamma\leftrightarrow \tau$, and $T\leftrightarrow D/\tau$. The final term captures the fundamental differences between AOUPs and {UBP}s, which results from the different ways pairwise forces enter the dynamics. Interestingly, this difference impacts fundamental properties such as the existence of MIPS or lack thereof, but not the absence of {directed} currents.

\subsection{Summary}
Despite their apparent similarities, the contrast between ABPs/RTPs and AOUPs is striking when it comes to interaction-induced {directed} currents. While ABPs and RTPs display interaction-induced {transport} in activity landscapes (Fig.~\ref{fig:abp-rtp-2d-phenomenology}), this is impossible for AOUPs. This is because, surprisingly, AOUPs with spatially varying activity admit a generalized stress tensor.

{In this article, we have focused on active particles with uniform persistence times and spatially-varying activity. In Eq.~\eqref{eq:langevin-aoup-b}, this amounts to setting $\tau$ constant and allowing $D$ to vary. Alternatively, one could study inhomogeneous $\tau(\br)$ with constant $D$, which breaks the existence of a stress tensor and can induce currents. This implies that the small-persistence-time limits of these models, which all lead to {OBP}s with inhomogeneous temperatures, correspond to different time discretizations for the noise. It would be interesting to study the small-persistence-time limits of these distinct flavours of AOUPs in more detail, for instance using the methods introduced in~\cite{bo_white-noise_2013}.}

\begin{figure*}
    \centering
    \includegraphics[width=7in]{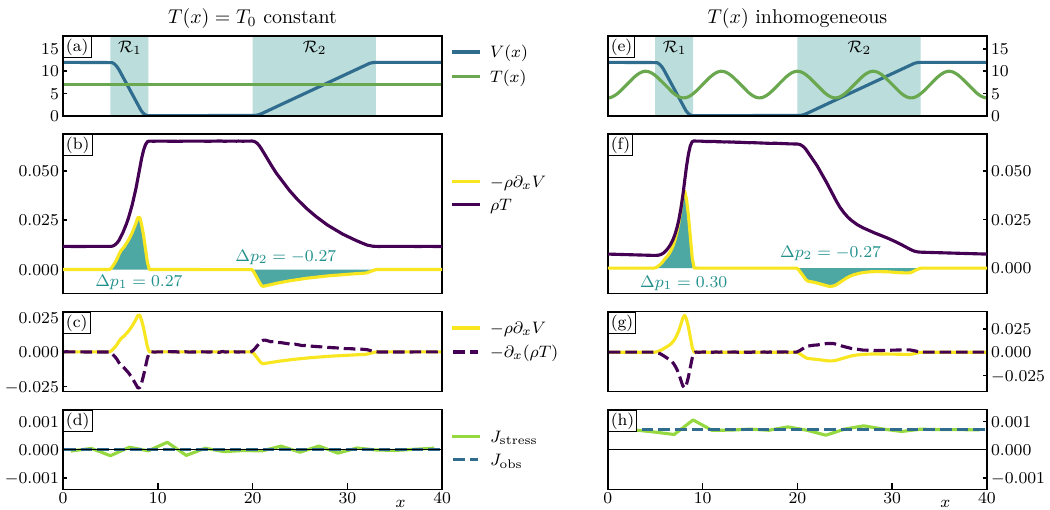}
    \caption{{\textbf{Momentum sources in the ``blowtorch ratchet."} Simulations of interacting {OBP}s in an external potential with uniform temperature [panels (a-d)] and with spatially-varying temperature [panels (e-h)]. 
    {\bf (a), (e)} Temperature and external potential used in the simulations. {\bf (b), (f)} Thermal stress $\rho(x) T(x)$ (dark purple) and momentum source $-\rho(x) \partial_x V(x)$ (yellow) measured in simulations. The momentum exchanged with the potential is shaded in blue. The two contributions $\Delta p_1$ and $\Delta p_2$ do not balance only when $T(x)$ is spatially-varying.
    {\bf (c), (g)} The contributions to the current from the stress (dashed purple) and from the momentum source (yellow). {\bf (d), (h)} $J_{\rm stress} = -\rho \partial_x V - \partial_x (\rho T)$, the current due to stress and momentum sources (dashed teal), agrees with the current $J_{\rm obs}$ measured from particle displacements. Simulation details are given in Appendix~\ref{appendix:sim-blowtorch}.}}
    \label{fig:blowtorch}
\end{figure*}

{
\section{Blowtorch ratchets: when fluctuation landscapes rectify momentum sources}}\label{sec:blowtorch}
{Until now, all calculations have been done in the absence of an external potential. In this section, we lift this restriction, and consider the consequences of introducing external forces into an inhomogeneous fluctuation landscape. We consider non-interacting {OBP}s which evolve according to the Langevin dynamics}
\begin{align}
    {\dot{\br}= -\nabla V(\br) + \sqrt{2 T(\br)} \bfeta(t)}
\end{align}
{with external potential $V(\br)$ and spatially-varying temperature $T(\br)$.

The rectification of {OBP}s by a combination of external forces and nonuniform temperature has already received considerable attention in the study of so-called ``blowtorch" ratchets---by locally increasing the temperature on one side of a potential, one can induce rectification of the particle motion over the potential~}\cite{buttiker_transport_1987,blanter_rectification_1998,landauer_motion_1988}{. 

Let us now study this system from the point of view of momentum conservation. First, we note that an external potential always acts as a momentum source, which cannot be absorbed into a stress. However, this does not mean that there is a net injection of momentum into the system. As a simple illustration, consider non-interacting {OBP}s with spatially-varying temperature in an external potential $V$, whose density evolves as}
\begin{equation}
    {\dot{\rho} = -\nabla \cdot \bJ\;,\quad \text{where}\quad\bJ = -\rho \nabla V - \nabla \cdot [\mathbb{I} \rho T]\;.}
\end{equation}
{When the temperature is constant, the system is in equilibrium, and the momentum sources $-\rho \nabla V$ integrates to zero since}
\begin{equation}
    {\int d\br \rho \nabla V\propto \int d \br  e^{-\beta V}\nabla V \propto \int d \br \nabla e^{-\beta V}=0\;,}
\end{equation} 
{where the last equality stems from the divergence theorem. The lack of {directed} current in this equilibrium system can thus be seen as a consequence of momentum sources balancing out---in addition to being enforced by time-reversal symmetry. This is illustrated in panels (a-d) of Fig.~\ref{fig:blowtorch}.}

{On the other hand, if the temperature is spatially-varying, the system exhibits a non-Boltzmann steady state and nothing prevents the existence of a net momentum source. In this case, the steady {directed} current can be seen as emerging from the imbalance of the momentum sources due to the external potential. This is demonstrated in panels (e-h) of Fig.~\ref{fig:blowtorch}, where {OBP}s are simulated in the presence of an external potential and a spatially-varying temperature. The temperature landscape cannot act as a momentum source, but it rectifies the forces due to the external potential, hence powering a {directed} current. Similar cases with symmetric potential and asymmetric temperature sources, or symmetric potential and temperature sources with a phase shift, show consistent results, detailed in Appendix~\ref{app:blowtorch} for completeness. Thus, the existence of a current still hinges on the existence of net momentum sources, in addition to the violation of TRS and parity symmetry.}

\section{Discussion and Conclusion}\label{sec:discussion-conclusion}

In this work, we have considered a series of stochastic particle systems for which we have shown that the emergence of {directed} currents requires the systematic violation of TRS {in position space}, parity symmetry, and bulk momentum conservation. 

By considering overdamped and underdamped {passive} Brownian particles as well as ABPs, RTPs, and AOUPs, we have shown that the litmus test for the emergence of interaction-induced {directed} currents in fluctuation landscapes is not activity. The existence of an effective momentum conservation law is thus a new dividing line between non-equilibrium systems that raises new questions. In particular, we have here focused on continuous diffusion models in space and the question as to how our results extend to more general Markov processes, multi-state systems, or lattice gases is open. The notion of effective momentum conservation does not easily generalize to such cases and the identification of the possible conservation laws that may prevent the emergence of {directed} currents in these contexts is an exciting open question for future work.

\textit{Acknowledgements.} We thank {Hugues Chate, Cory Hargus, }Yariv Kafri, Jeremy O'Byrne, and Vivien Lecomte for insightful discussions.

\appendix

\section{Simulation details}
\label{app:sim}
All simulations are run with periodic boundary conditions in all directions, using a standard Euler(-Maruyama) discretization scheme. All interacting simulations were done using the soft repulsive harmonic potential and force given by:
\begin{align}
    \label{eq:appendix-Uint}\U(\br) &= \frac{\varepsilon r_{\rm int}}{2} \left[1 - \frac{|\br|}{r_{\rm int}}\right]^2\equiv \varepsilon \Tilde{U}_{\rm int}(\br)\;,\\
    \mathbf{f}_{\rm int}(\br) &= -\nabla \U(\br) = \varepsilon \frac{\br}{|\br|} \left[1 - \frac{|\br|}{r_{\rm int}}\right]\;.
\end{align}
Our unit of length are such that the interaction range is $r_{\rm int}=1$. For the ABP, RTP, and AOUP simulations, time units are such that $\tau=1$ (with $\tau=\alpha^{-1}$ for RTPs and $\tau=D_r^{-1}$ for ABPs). For {OBP and UBP} simulations, the mobility or inverse damping is similarly set to $\mu=\gamma^{-1}=1$. 

All activity, diffusivity, and temperature landscapes presented in this paper, with the exception of those used in Figs.~\ref{fig:ratchet-examp},~\ref{fig:1d-rtp-interact-profile-disp}, and~\ref{fig:RTP-U-v-exact}, are constructed using cubic interpolation between some minimum and maximum values. More precisely, consider a landscape $f(x)$ ranging from $f_0$ to $f_1$ using reference points $x_\ell< x_{c1}\leq x_{c2}< x_r$. Define the interpolation distances $\Delta x_1 \equiv x_{c1}-x_\ell$ and $\Delta x_2 \equiv x_r-x_{c2}$, and the length of the central segment $\Delta x_c \equiv x_{c2}-x_{c1}$. Then $f$ has the functional form 
\begin{align}
    f(x) &= \begin{cases}
        f_1,\quad &x\in [0,x_\ell]\cup [x_r,L_x]\\
        f_0,\quad &x\in [x_{c1},x_{c2}]\\
        f_1 + (f_0-f_1) s\left( \frac{x-x_\ell }{\Delta x_1}\right),\quad &x \in [x_\ell,x_{c1}]\\
        f_0 + (f_1-f_0)s\left( \frac{x-x_{c2} }{\Delta x_2}\right)\quad &x\in[x_{c2},x_r]\;,
    \end{cases}
    \label{eq:fluct-landscape-func}\\
    s(x) &\equiv -2x^3 + 3x^2\;.
\end{align}

Note that while our analytic calculations often use the normalization $\int d\br \rho(\br)=N$, our figures display simulation data that is normalized such that $\int d\br \rho(\br)=1$, for ease of comparison with non-interacting systems of $N=1$.

\begin{table*}[]
\renewcommand{\arraystretch}{1.65} 
    \centering
    \begin{tabular}{c|c|c|c|c|c|c|c|c|c|c|c|c|c}
        Fig. & $N$ & $L_x$ & $L_y$ & $f_0$ & $f_1$ & $\Delta x_1$ & $\Delta x_c$ & $\Delta x_2$ & $\varepsilon$ & dt & $\tf$ & $n_{\rm seed}$ & notes\\
        \hline
        \ref{fig:ratchet-examp} & {120} & {30} & {10} & {1} & {2.5} & {2} & {0} & {8} & {\{0,10\}} & {$2.5 \times 10^{-4}$} & {$8 \times 10^6$} & {15} & {$f(x)$ repeated 3$\times$}
        \\
        \ref{fig:ud-pbp-eos} & 700 & 60 & 15 & 20 & 1 & 10 & 10 & 30 & $\{0,50\}$ & $2.5\times 10^{-4}$ & $2\times 10^4$ & 30\\
        \ref{fig:od-pbp-disc-induced-current} & 200 & 20 & 5 & 20 & 1 & 2.5 & 0 & 17.5 & 50 & $2.5\times 10^{-4}$ & $2\times 10^5$ & 200 & $\alpha \in \{0, 0.125, \ldots, 1\}$\\
        \ref{fig:EOSpassive} & 400 & 50 & 15 & 20 & 1 & 8 & 9 & 24 & 50 & $5\times 10^{-4}$ & $10^5$ & 1 &\\
        \ref{fig:abp-rtp-2d-phenomenology} & 500 & 20 & 5 & 20 & 1 & 2.5 & 2.5 & 7 & $\{0,0.025,\ldots,0.225\}$ & $5 \times 10^{-4}$ & $2\times 10^4$ & 400 &\\
        \ref{fig:1d-rtp-interact-profile-disp} & $\{12,25,\ldots,400\}$ & 10 & - & - & - & - & - & - & $15/N$ & $5 \times 10^{-4}$ & varies & varies & 1d sinusoidal $v(x)$\\
        \ref{fig:aoup-eos} & 700 & 60 & 15 & 20 & 1 & 10 & 10 & 30 & $\{0,50\}$ & $2.5\times 10^{-4}$ & $2\times 10^4$ & 30\\
        \ref{fig:blowtorch} & {1} & {40} & {15} & {4} & {10} & {4} & {0} & {4} & {0} & {$10^{-4}$} & {$10^5$} & {3,200}& {$V(x)$ details in text}\\
        \ref{fig:RTP-U-v-exact} & 1 & 20 & - & - & - & - & - & - & 0 & $10^{-3}$ & $10^3$ & 96,000 & 1d sinusoidal $v(x),\,U(x)$\\
        \ref{fig:blowtorch-app} & {1} & {40} & {15} & {4} & {10} & {4} & {0} & {4} & {0} & {$10^{-4}$} & {$10^5$} & {1,600} & {panels (a)-(d)}\\
        \ref{fig:blowtorch-app} & {1} & {40} & {15} & {4} & {10} & {4} & {16} & {4} & {0} & {$10^{-4}$} & {$10^5$} & {1,600} & {panels (e)-(h)}\\
        \ref{fig:blowtorch-app} & {1} & {40} & {15} & {4} & {10} & {6} & {0} & {2} & {0} & {$10^{-4}$} & {$10^5$} & {1,600} & {panels (i)-(l)}
    \end{tabular}
    \caption{Simulation parameters for each figure. All {simulations} have translational symmetry in all but the $x$ dimension. All but Figs.~\ref{fig:1d-rtp-interact-profile-disp} and~\ref{fig:RTP-U-v-exact} are in 2 space dimensions. All but Figs.~\ref{fig:1d-rtp-interact-profile-disp} and~\ref{fig:RTP-U-v-exact} use a fluctuation landscape $f(x)$ that interpolates cubically between the values $f_0$ to $f_1$ [Eq.~\eqref{eq:fluct-landscape-func}]. {The temperature landscape is repeated 5 times in Figs.~\ref{fig:blowtorch}(e)-(h), Fig.~\ref{fig:blowtorch-app}(a)-(d), and Fig.~\ref{fig:blowtorch-app}(i)-(l). In Figs.~\ref{fig:blowtorch} and~\ref{fig:blowtorch-app}, the external potential $V(x)$ is detailed in Sec.~\ref{appendix:sim-blowtorch}.}}
    \label{tab:sim-params}
\end{table*}

Below, we describe the simulations presented in the figures in the main text. A summary of these parameters is given in Table~\ref{tab:sim-params}.

\subsection{Fig.~\ref{fig:ratchet-examp}}\label{appendix:sim-examp}

{The simulations shown in Fig.~\ref{fig:ratchet-examp} were performed using $N=120$ particles ({UBP}s, {OBP}s, ABPs, and AOUPs as indicated in headings) with interaction strength $\varepsilon=10$. The domain has size $L_x=30$ and $L_y=10$. The fluctuation landscape $T(x)$, $v(x)$, or $D(x)$ is constructed as in Eq.~\eqref{eq:fluct-landscape-func} with reference points $x_\ell=0,\,x_{c1}=x_{c2}=2,\,x_r=10$; and repeated thrice.}

\subsection{Underdamped Passive Brownian Particles (Fig.~\ref{fig:ud-pbp-eos})}\label{appendix:sim-ud-pbp}
The simulations shown in Fig.~\ref{fig:ud-pbp-eos} were performed using $N=700$ particles with interaction strength $\varepsilon=0$ (gold lines) and $\varepsilon=50$ (dark red lines). The domain has size $L_x=60$ and $L_y=15$. The temperature ranges from $T_0=1$ to $T_1=20$. The temperature field is constructed as in Eq.~\eqref{eq:fluct-landscape-func} with reference points $x_\ell=0,\,x_{c1}=10,\,x_{c2}=20,\,x_r=50$.

\subsection{Overdamped Passive Brownian Particles (Figs.~\ref{fig:od-pbp-disc-induced-current}-\ref{fig:EOSpassive})}\label{appendix:sim-od-pbp}
The simulations shown in Fig.~\ref{fig:od-pbp-disc-induced-current} were performed using $N=200$ particles with interaction strength $\varepsilon=50$. The domain has size $L_x=20$ and $L_y=5$. The temperature ranges from $T_0=1$ to $T_1=20$. The temperature field is constructed as in Eq.~\eqref{eq:fluct-landscape-func} with reference points $x_\ell=0$, $x_{c1}=x_{c2}=2.5$, $x_r=L_x$.

Simulations of {OBP}s with discretization $\alpha\neq 0$ were performed with the same Euler discretization, which corresponds to an It\=o process, and varying the corresponding spurious drift~\cite{gardiner2009stochastic}:
\begin{align}
    \dot{\br}_i &\overset{\alpha}{=} \bF_i(\{\br_j\}) + \sqrt{2 T(\br_i)} \bfeta_i\label{eq:appendix-discretization-shortcut-1}\\
    \Leftrightarrow \dot{\br}_i &\overset{0}{=} \bF_i(\{\br_j\}) + \alpha \nabla T(\br_i) + \sqrt{2 T(\br_i)} \bfeta_i\;.\label{eq:appendix-discretization-shortcut-2}
\end{align}

The simulations shown in Fig.~\ref{fig:EOSpassive} were performed using $N=400$ particles with interaction strength $\varepsilon=50$. The domain has size $L_x=50$ and $L_y=15$. The temperature ranges from $T_0=1$ to $T_1=10$ with the reference points of Eq.~\eqref{eq:fluct-landscape-func} given by $x_\ell=0,\,x_{c1}=8,\,x_{c2}=17,\,x_r=41$. The EPR profile was measured using $2 \times 10^7$ equally-spaced time samples.

\subsection{Active Brownian and Run-and-Tumble Particles (Figs.~\ref{fig:abp-rtp-2d-phenomenology}-\ref{fig:1d-rtp-interact-profile-disp},~\ref{fig:RTP-U-v-exact})}\label{appendix:sim-abp-rtp}
The simulations shown in Fig.~\ref{fig:abp-rtp-2d-phenomenology} were each performed using $N=500$ particles with interaction strengths $\varepsilon \in \{0,0.025,0.05,\ldots, 0.225\}$. The domain has size $L_x=20$ and $L_y=5$. The activity ranged from $v_0=1$ to $v_1=20$ with the reference points of Eq.~\eqref{eq:fluct-landscape-func} given by $x_\ell=0,\,x_{c1}=2.5,\,x_{c2}=5,\,x_r=12$.

The simulations shown in Fig.~\ref{fig:1d-rtp-interact-profile-disp} involved interacting 1d RTPs with sinusoidal activity profile $v(x)= 20\left[0.5 \sin\big(\frac{2\pi x}{L_x}\big) + 0.4 \sin\big(\frac{4\pi x}{L_x}\big)\right]$. Different simulations were run with different numbers of particles $N \in \{12,25,50,100,200,400\}$. The interaction strengths $\varepsilon$ are chosen such that $N\varepsilon=15$. The simulation lengths $t_{\rm f}$ and number of seeds $n_{\rm seed}$ vary with the number of particles $N$ as indicated in the following list of tuples $(N,t_{\rm f},n_{\rm seed})$: $(12,6\times 10^7,480)$, $(25,6\times 10^7,240)$, $(50, 5\times 10^7,400)$, $(100,8\times 10^6,240)$, $(200,4\times 10^6,360)$, and $(400,2\times 10^6,400)$.

The simulations shown in Fig.~\ref{fig:RTP-U-v-exact} involved 96,000 independent RTPs in a 1-dimensional domain of size $L=20$. The activity and potential landscapes $v(x)$ and $U(x)$ were constructed as the sum of 12 Fourier modes, 
\begin{align}
    v(x) =& \sum_{n=1}^{12} a_n \sin(2\pi (n x/L + \phi_n))\\
    U(x) =& \sum_{n=1}^{12} b_n \sin(2\pi (nx/L + \psi_n))\;.
\end{align}
They ran until the time $\tf=1000$, with a timestep dt$=10^{-3}$.

\subsection{Active Ornstein Uhlenbeck Particles (Fig.~\ref{fig:aoup-eos})}\label{appendix:sim-aoup}
The simulations shown in Fig.~\ref{fig:aoup-eos} were performed using $N=700$ particles with interaction strength $\varepsilon=0$ (gold lines) and $\varepsilon=50$ (dark red lines). The domain has size $L_x=60$ and $L_y=15$. The activity ranges from $D_0=1$ to $D_1=20$. The activity field is constructed as in Eq.~\eqref{eq:fluct-landscape-func} with reference points $x_\ell=0,\,x_{c1}=10,\,x_{c2}=20,\,x_r=50$.

{
\subsection{Blowtorch ratchets (Figs.~\ref{fig:blowtorch} and~\ref{fig:blowtorch-app})}\label{appendix:sim-blowtorch}
The simulations shown in Figs.~\ref{fig:blowtorch} and~\ref{fig:blowtorch-app} each involved 6,400 non-interacting {OBP}s in a temperature landscape $T(x)$ and external potential $V(x)$. 

In Fig.~\ref{fig:blowtorch} panels (e)-(h), the temperature field is constructed by repeating Eq.~\eqref{eq:fluct-landscape-func} 10 times, with $x_\ell=0,\,x_{c1}=x_{c2}=4,\,x_{r}=8$ and $f_0=4$ and $f_1=10$. In both panels (a)-(d) and (e)-(h), the external force $F(x)$ is constructed by linearly interpolating between the values $0,0,4,4,0,0,-1,-1,0,0$ at $x=0,5,6,8,9,19,20,34,35,40$ respectively.

In Fig.~\ref{fig:blowtorch-app} panels (a)-(d), the temperature field is constructed in the same way as in Fig.~\ref{fig:blowtorch}(e)-(h). The force landscape is constructed by linearly interpolating between the values 4,-4,4,-4,4,-4,4,-4,4,-4 at $x=0,4,8,12,16,20,24,28,32,36$ respectively.

In Fig.~\ref{fig:blowtorch-app} panels (e)-(h), the temperature field is given by Eq.~\eqref{eq:fluct-landscape-func} with $x_\ell=8$, $x_{c1}=12$, $x_{c2}=28$, $x_r=32$, and $f_0=10$ and $f_1=4$. The force landscape is constructed the same way as in panels (a)-(d).

In Fig.~\ref{fig:blowtorch-app} panels (i)-(l), the temperature field is constructed by repeating Eq.~\eqref{eq:fluct-landscape-func} 10 times, with $x_\ell=0$,\, $x_{c1}=x_{c2}=6,\,x_r=8,$ and $f_0=4$ and $f_1=10$. The external force $F(x)$ is constructed by linearly interpolating between the values 0,4,4,0,0,-4,-4,0 at $x=7,8,12,13,27,28,32,33$ respectively.

The external potential $V(x)$ is such that $F(x) = -\partial_x V(x)$.

\section{Derivation of the Irving-Kirkwood stress tensor}\label{app:irving-kirkwood}

Here we provide a derivation of the Irving-Kirkwood stress tensor $\bsigma_\subIK$~\cite{irving1950statistical}. The contribution to the current density at position $\br$ due to interparticle forces reads $ \sum_{ij} \bF_{ij}\delta(\br -\br_i)$, where $\bF_{ij} = -\nabla \U(\br_i-\br_j)$. Any conservative force is  reciprocal, so
\begin{align}
 \sum_{ij} \bF_{ij}\delta(\br -\br_i)
 &= \frac{1}{2} \sum_{ij} (\bF_{ij} - \bF_{ji}) \delta(\br -\br_i)\\
 &= \frac{1}{2} \sum_{ij} \bF_{ij}\big[\delta(\br -\br_i) - \delta(\br -\br_j) \big]\;,\label{eq:Irving-Kirkwood}
 \end{align}
 where the second equality stems from reindexing. Introducing $\br_{ij} \equiv \br_i - \br_j$, the chain rule implies
 \begin{equation}
     \partial_\lambda \delta(\br - \br_j - \lambda\br_{ij})= - \br_{ij} \cdot  \nabla \delta(\br - \br_j - \lambda\br_{ij}) \;.
 \end{equation}
Integrating both sides from $\lambda=0$ to $\lambda=1$ then leads to
\begin{equation*}
    \delta(\br -\br_i) - \delta(\br -\br_j)
    = -\nabla \cdot \bigg[\br_{ij} \int_0^1 d\lambda\ \delta(\br - \br_j - \lambda\br_{ij}) \bigg]\,.
\end{equation*}
In turn, this allows rewriting Eq.~\eqref{eq:Irving-Kirkwood} as
\begin{align}
 \sum_{ij} \bF_{ij}\delta(\br -\br_i)
 &=   - \nabla \cdot \sum_{ij} \frac{\br_{ij}}2 \otimes \bF_{ij} \int_0^1 d\lambda\ \delta(\br - \br_j - \lambda \br_{ij}) \notag\\
 &\equiv \nabla\cdot \bsigma_\subIK\,,
\end{align}
where the last line is the definition of $\bsigma_\subIK$:
\begin{equation}
\bsigma_\subIK=- \sum_{ij} \frac{\br_{ij}}2 \otimes \bF_{ij} \int_0^1 d\lambda\ \delta(\br - \br_j - \lambda \br_{ij})\;.
\end{equation}

}

\section{EPR for interacting {UBP}s in temperature fields}\label{appendix:path-integ-upbp-int}
Here we provide a detailed calculation of the EPR given in Eq.~\eqref{eq:ud-pbp-epr-density} for interacting {UBP}s in an inhomogeneous temperature field. We calculate this by directly computing the probability of a trajectory with dynamics defined in Eq.~\eqref{eq:langevin-ud-pbp}, conditioned on the particles' starting positions and momenta. While no discretization ambiguity afflicts these dynamics, treating the discretization carefully is necessary when calculating path probabilities. Thus, we define the discretized timesteps $t_k=k \Delta t$ for $k\in \{0,1,\ldots,M\}$ where $M\Delta t = \tf$. We also consider a Stratonovich-discretized version of the dynamics in position space and thus introduce the position at the midpoint of the timestep $k$,
\begin{align}
    \overline{\br}_i^k &\equiv \frac{\br_i^k + \br_i^{k+1}}{2}\;,\label{eq:overline-modpoint-strato-definition}
\end{align}
and the spatial increments
\begin{align}
    \Delta \br_i^k &\equiv \br_i^{k+1} - \br_i^k\;.\label{eq:spatial-increment-deltar}
\end{align}
Starting from the continuous-time dynamics
\begin{equation}
        \ddot{\br}_i = -\gamma \dot{\br}_i + \sum_{j=1}^N \bfint(\br_i-\br_j) + \sqrt{2\gamma T(\br_i)} \bfeta_i\;,
\end{equation}
the discretized dynamics reads
\begin{align}
    \frac{\Delta \br_i^{k+1} - \Delta \br_i^{k}}{\Delta t^2} &= -\gamma \left(\frac{\Delta \br_i^k}{\Delta t}\right)+ \sum_{j=1}^N  \bfint(\overline{\br}_i^k - \overline{\br}_j^k)\notag \\
    &\quad  + \sqrt{2\gamma T(\overline{\br}_i^k)} \Delta \bfeta_i^{k+1}\;,\label{eq:ud-pbp-discretized-langevin}
\end{align}
where  $\Delta \bfeta_i^{k+1}$ is a centered Gaussian noise that satisfies $\langle \Delta \eta_{i,\mu}^k \Delta \eta_{j,\nu}^\ell\rangle = \delta_{ij} \delta_{\mu\nu} \delta_{k,\ell} /\Delta t$. 
We know the probability of a noise trajectory,
\begin{equation}
    \mathbb{P}[\{\Delta \bfeta_i^k\}\big|\{\br_i^0,\br_i^1\}] \propto \exp\left[-\sum_{i=1}^N \sum_{k=1}^{M-1} \frac{\Delta t}{2} |\Delta \bfeta_i^k|^2\right]\;,\label{eq:ud-pbp-noise-prob}
\end{equation}
from which the probabilities of the spatial trajectories can be found using the equality
\begin{align}\label{eq:probatrajupbp}
    \mathbb{P}[\{\br_i^k\}\big|\{\br_i^0,\br_i^1\}] &= \mathbb{P}[\{\Delta \bfeta_i^k\}\big|\{\br_i^0,\br_i^1\}] \left|\frac{\mathbb{D}[\{\Delta \bfeta_i^k\}\big|\{\br_i^0,\br_i^1\}]}{\mathbb{D}[\{\br_i^k\}\big|\{\br_i^0,\br_i^1\}]}\right|
\end{align}
where $\big|{\mathbb{D}[\{\Delta \bfeta_i^k\}\big|\{\br_i^0,\br_i^1\}]}/{\mathbb{D}[\{\br_i^k\}\big|\{\br_i^0,\br_i^1\}]}\big|$ is the Jacobian between the $M-1$ variables $\{\Delta \bfeta_i^k\}_{k=1}^{M-1}$ and the $M-1$ variables $\{\br_i^k\}_{k=2}^M$. Note the shift in $k$ indices: $\Delta \bfeta_i^k$ determines $\br_i^{k+1}$. Note that we condition on the positions in the first two timesteps, $\{\br_i^0,\br_i^1\}$, which is equivalent to conditioning on the initial positions and momenta. We first note, using the relation between the variables given by Eq.~\eqref{eq:ud-pbp-discretized-langevin}, that this Jacobian will be upper triangular in the time dimension, because 
\begin{align}
    \pfrac{\Delta \eta_{i,\mu}^{k+1}}{r_{j,\nu}^\ell} = 0 \;\;\text{whenever}\;\;\ell>k+2\;.
\end{align}
For the diagonal elements $\partial \Delta \eta_{i,\mu}^{k+1}/\partial r_{j,\nu}^{k+2}$, we have
\begin{align}
    \frac{\delta_{ij} \delta_{\mu\nu}}{\Delta t^2} &= \sqrt{2\gamma T(\overline{\br}_i^k)} \pfrac{\Delta \eta_{i,\mu}^{k+1}}{ r_{j,\nu}^{k+2}}
\end{align}
from which we find the determinant
\begin{align}
    \left|\frac{\mathbb{D}[\{\Delta \bfeta_i^k\}\big|\{\br_i^0,\br_i^1\}]}{\mathbb{D}[\{\br_i^k\}\big|\{\br_i^0,\br_i^1\}]}\right| &= \prod_{k=0}^{M-2} \prod_{i=1}^N \frac{1}{\Delta t^2 \sqrt{2\gamma T(\overline{\br}_i^k)}} \label{eq:ud-pbp-jac-det}\\
    &\propto \prod_{k=0}^{M-2} \prod_{i=1}^N \frac{1}{\sqrt{T(\overline{\br}_i^k)}}\;.
\end{align}
Next, we convert the noise probability [Eq.~\eqref{eq:ud-pbp-noise-prob}] into spatial coordinates using Eq.~\eqref{eq:ud-pbp-discretized-langevin}:
\begin{align}\label{eq:pbp-trajfinal}
    \mathbb{P}&[\{\Delta \bfeta_i^k\}\big|\{\br_i^0,\br_i^1\}] \\
    &\propto \exp\bigg[-\sum_{i=1}^N \sum_{k=0}^{M-2} \frac{\Delta t }{4\gamma T(\overline{\br}_i^k)} \bigg|\frac{\Delta \br_i^{k+1}-\Delta \br_i^k}{\Delta t^2} \nonumber\\
    &\qquad + \gamma \frac{\Delta \br_i^k}{\Delta t} - \sum_{j=1}^N \bfint(\overline{\br}_i^k - \overline{\br}_j^k)\bigg|^2\bigg]\nonumber\;.
\end{align}
Equations~\eqref{eq:probatrajupbp},~\eqref{eq:ud-pbp-jac-det}, and~\eqref{eq:pbp-trajfinal} give the trajectory probability density.

We wish to compare the probability of observing a trajectory to the probability of observing its time-reversed counterpart, to isolate the ``time-irreversible parts" of the dynamics. We use the definition a trajectory's time-reverse $\br_i^R$
\begin{align}
    \br_i^{R,k} &\equiv \br_i^{M-k}\;,
\end{align}
which satisfies $\overline{\br}_i^{R,k}=\overline{\br}_i^{M-k-1}$ and $\Delta \br_i^{R,k}=-\Delta \br_i^{M-k-1}$. We find the ratio between the probabilities of observing a trajectory and its reverse as:
\begin{widetext}
\begin{align}
    \frac{\mathbb{P}[\{\br_i^k\}\big|\{\br_i^0,\br_i^1\}]}{\mathbb{P}[\{\br_i^{R,k}\}\big|\{\br_i^{R,0},\br_i^{R,1}\}]} =& \sqrt{\frac{T(\overline{\br}_i^{M-1})T(\overline{\br}_i^{M-2})}{T(\overline{\br}_i^0)T(\overline{\br}_i^1)}}
     \exp\bigg[-\sum_{i=1}^N \sum_{k=0}^{M-2} \bigg\{\frac{\Delta t }{4 \gamma T(\overline{\br}_i^k)}  \bigg|\frac{\Delta \br_i^{k+1}-\Delta \br_i^k}{\Delta t^2} + \gamma \frac{\Delta \br_i^k}{\Delta t} - \sum_{j=1}^N \bfint(\overline{\br}_i^k - \overline{\br}_j^k)\bigg|^2\nonumber\\
    & - \frac{\Delta t }{4 \gamma T(\overline{\br}_i^{M-k-1})} \bigg|\frac{-\Delta \br_i^{M-k-2}+\Delta \br_i^{M-k-1}}{\Delta t^2} 
     - \gamma \frac{\Delta \br_i^{M-k-1}}{\Delta t} - \sum_{j=1}^N \bfint(\overline{\br}_i^{M-k-1} - \overline{\br}_j^{M-k-1})\bigg|^2\bigg\}\bigg]\nonumber\\
    =& \sqrt{\frac{T(\overline{\br}_i^{M-1})T(\overline{\br}_i^{M-2})}{T(\overline{\br}_i^0)T(\overline{\br}_i^1)}} \exp\bigg[-\sum_{i=1}^N \sum_{k=0}^{M-2} \frac{\Delta t }{4 \gamma T(\overline{\br}_i^k)}  \bigg|\frac{\Delta \br_i^{k+1}-\Delta \br_i^k}{\Delta t^2} + \gamma \frac{\Delta \br_i^k}{\Delta t} - \sum_{j=1}^N \bfint(\overline{\br}_i^k - \overline{\br}_j^k)\bigg|^2\nonumber\\
    &\: + \sum_{i=1}^N \sum_{k=1}^{M-1}  \frac{\Delta t }{4 \gamma T(\overline{\br}_i^k)} \cdot \bigg|\frac{\Delta \br_i^k-\Delta \br_i^{k-1}}{\Delta t^2} - \gamma \frac{\Delta \br_i^k}{\Delta t} -\sum_{j=1}^N \bfint(\overline{\br}_i^k - \overline{\br}_j^k)\bigg|^2\bigg]\;,\label{eq:noname}
\end{align}
\end{widetext}
where, in the second equality, we have relabelled the time indices for the backward path as $M-k-1 \to k$. In the last line of Eq.~\eqref{eq:noname}, we may replace $\frac{\Delta \br_i^k-\Delta \br_i^{k-1}}{\Delta t^2}$ with $\frac{\Delta \br_i^{k+1}-\Delta \br_i^k}{\Delta t^2}$, since their difference is of order $\mathcal{O}(\sqrt{\Delta t})$ and thus do not contribute to the sum in the limit $\Delta t \to 0$. When expanding the squared terms, many cancelations occur. Taking the natural logarithm of Eq.~\eqref{eq:noname} then leads to the following result for the entropy production $\Sigma$ along the path:
\begin{widetext}
\begin{align}
    \Sigma(\tf) &\equiv \ln \bigg[\frac{\mathbb{P}[\{\br_i^k\}\big|\{\br_i^0,\br_i^1\}]}{\mathbb{P}[\{\br_i^{R,k}\}\big|\{\br_i^{R,0},\br_i^{R,1}\}]}\bigg]\\
    &=\frac12 \ln \bigg[\frac{T(\overline{\br}_i^{M-1})T(\overline{\br}_i^{M-2})}{T(\overline{\br}_i^0)T(\overline{\br}_i^1)}\bigg]
    \;-\bigg\{ \sum_{i=1}^N \sum_{k=1}^{M-2} \frac{\Delta t}{T(\overline{\br}_i^k)}  \frac{\Delta \br_i^k}{\Delta t} \cdot \bigg[\frac{\Delta \br_i^{k+1}-\Delta \br_i^k}{\Delta t^2}- \sum_{j=1}^N \bfint(\overline{\br}_i^k - \overline{\br}_j^k)\bigg]\bigg\}+\mathcal{O}(\sqrt{\Delta t})\label{eq:ud-pbp-boundaryterms}\\
    &\underset{\Delta t\rightarrow 0}{\longrightarrow} \ln \left[\frac{T(\overline{\br}_i(\tf))}{T(\overline{\br}_i(0))}\right] - \sum_{i=1}^N \int_0^\tf dt \frac{ \dot{\br}_i  }{T(\br_i)} \cdot \bigg[\ddot{\br}_i - \sum_{j=1}^N \bfint(\br_i-\br_j)\bigg]\;.
\end{align}   
\end{widetext}
In Eq.~\eqref{eq:ud-pbp-boundaryterms}, the $\mathcal{O}(\sqrt{\Delta t})$ term corresponds to the missing $k=0$ and unmatched $k=M-1$ terms that appear in the last line of Eq.~\eqref{eq:noname}. They do not contribute in the limit $\Delta t\to 0$.

The EPR $\sigma$ is then defined as $    \sigma = \lim_{\tf\rightarrow\infty} \frac{\Sigma(\tf)}{\tf}$ and we thus find:
\begin{align}
    \sigma = \sum_{i=1}^N \left\langle\frac{ \dot{\br}_i}{T(\br_i)} \cdot \bigg[\sum_{j=1}^N \bfint(\br_i-\br_j) - \ddot{\br}_i\bigg]\right\rangle\;,
\end{align}
where we have assumed the system to be ergodic. In Eq.~\eqref{eq:ud-pbp-epr-density}, we then define a fluctuating EPR density $\hat{\sigma}(\br)$ such that $\int d\br \langle \hat{\sigma}(\br)\rangle=\sigma$.

\section{EPR for {OBP}s}

\subsection{TRS of non-interacting {OBP}s}\label{appendix:path-integ-pbp-ni}
It is useful to convert the $\alpha$-discretized Langevin equation~\eqref{eq:od-pbp-alpha-langevin-disc} into an equivalent Stratonovich-discretized one. This incurs an extra force $\left(\alpha-\frac12\right) \nabla T(\br)$, and the new Langevin equation is
\begin{align}
    \dot{\br} &\overset{1/2}{=} \Big(\alpha-\frac12\Big) \nabla T({\br}) + \sqrt{2 T({\br})} \bfeta\;.
\end{align}
In discretized form, we may write this as
\begin{align}
    \Delta \br^k &= \Big(\alpha-\frac12\Big) \nabla T(\overline{\br}^k) \Delta t + \sqrt{2 T(\overline{\br}^k) } \Delta \bfeta^k\;.
    \label{eq:od-pbp-ni-alpha-to-strato}
\end{align}
where are using $\overline{\br}^k$, the position evaluated at the midpoint of the timestep, as defined in Eq.~\eqref{eq:overline-modpoint-strato-definition}. We index our timesteps by $k\in \{0,\ldots,M\}$, and take starting time $t_0=0$ and finising time $t_M\equiv \tf$.

We know the probability of a noise realization:
\begin{align}
    \mathbb{P}[\{\Delta \bfeta^k\}] &= \frac{1}{(2\pi)^{dN/2}} \exp\left[-\frac{1}{2\Delta t} \sum_{k=0}^{M-1} |\Delta \bfeta^k|^2\right]
    \label{eq:od-pbp-noise-prob}
\end{align}
and can convert this to a probability for the spatial trajectory using the Jacobian determinant of the associated transformation
\begin{align}
    \mathbb{P}[\{\br^k\}] &= \mathbb{P}[\{\Delta \bfeta^k\}] \bigg|\frac{\mathbb{D}[\{\Delta \bfeta^k\}]}{\mathbb{D}[\{\br^k\}]}\bigg| \\
    &= \mathbb{P}[\{\Delta \bfeta^k\}] \det_{\mu,\nu,k,\ell} \left(\pfrac{\Delta \eta^k_\mu}{r^\ell_\nu}\right)\;,
    \label{eq:od-pbp-traj-prob-conversion}
\end{align}
where $\mu,\nu\in\{1,\ldots,d\}$, $k\in \{0,\ldots,M-1\}$, and $\ell \in \{1,\ldots,M\}$. The quantity $\det_{\mu,\nu,k,\ell} \left({\partial \Delta \eta^k_\mu}/{\partial r^\ell_\nu}\right)$ can be understood as the determinant of the $Md\times Md$ matrix whose $(kd+\mu,\,\ell d+\nu)$ entry is ${\partial \Delta \eta^k_\mu}/{\partial r^\ell_\nu}$. 

Note first that $\Delta \bfeta^k$ only depends directly on $\br^k$ and $\br^{k+1}$ by the relation
\begin{align}
    \Delta \eta_\mu^k(\br^k,\br^{k+1}) &= \frac{\Delta r^k_\mu - \Delta t\big(\alpha-\frac12\big) \partial_\mu T(\overline{\br}^k)}{\sqrt{2 T(\overline{\br}^k) }}\;.
    \label{eq:od-pbp-langevin-noise-rearranged}
\end{align}
If we write the Jacobian matrix in block form, made of $M \times M$ blocks of $d \times d$ matrices, it is block triangular, i.e. there exist $d \times d$ matrices $A$ and $B$ such that
\begin{align}
    \pfrac{\Delta \eta_\mu^k}{r_\nu^\ell} = \delta^{k,\ell} A_{\mu\nu} + \delta^{k+1,\ell} B_{\mu\nu}\;.
\end{align}
Consequently, the determinant is given by the product of the determinants of the diagonal blocks. The diagonal blocks consist of $k,\ell$ such that $\ell=k+1$ (this is due to the fact that $k$ runs from 0 to $M-1$ while $\ell$ runs from 1 to $M$). Their elements are
\begin{align}
    \pfrac{\Delta \eta_\mu^k}{r_\nu^{k+1}} &= \frac{1}{\sqrt{2 T }} \bigg\{ \delta_{\mu\nu}  - \frac{\Delta r^k_\mu \partial_\nu T}{4 T}\\
    &\quad + \frac{\Delta t\big(\alpha-\frac12\big) }{2} \left[ \frac{\partial_\mu T \partial_\nu T}{2T} -\partial_\mu \partial_\nu T \right] \bigg\}\;,\nonumber
\end{align}
where each occurrence of $T$ is evaluated at $\overline{\br}^k$. Note that the diagonal elements ($\mu=\nu$) are $\mathcal{O}(1)$, while off-diagonal elements are $\mathcal{O}(\Delta t^{1/2})$. The off-diagonal elements are thus subleading to diagonal elements, and the determinant is just the product of diagonal elements to leading order. This yields
\begin{align}
    \det_{\mu,\nu} \left[\pfrac{\Delta \eta_\mu^k}{r_\nu^{k+1}}\right] &= \frac{1}{(2 T)^{d/2}} \bigg\{1 - \frac{\Delta \br^k \cdot \nabla T}{4T} \\
    &\quad+ \frac{\Delta t (\alpha-\frac12)}{2} \left[\frac{|\nabla T|^2}{2T} - \nabla^2 T\right]\bigg\}\;.\nonumber
\end{align}
Taking the product over $k$ yields the determinant of the entire matrix $\frac{\mathbb{D}[\{\Delta \bfeta^k\}]}{\mathbb{D}[\{\br^k\}]}$. Executing this and converting the product into an exponentiated integral in the $M\rightarrow\infty$ and $\Delta t\rightarrow 0$ limit, we find
\begin{align}
    \bigg|\frac{\mathbb{D}[ \bfeta(t)]}{\mathbb{D}[\br(t)]}\bigg| &= \bigg(\prod_{k=0}^{M-1} \frac{1}{(2 T(\overline{\br}^k))^{d/2}}\bigg)\label{eq:od-pbp-jacobian-result}\\
    &\quad \times \exp\bigg[ \int_0^{\tf} dt \bigg\{ - \frac{\dot{\br} \cdot \nabla T({\br})}{4T({\br})} \nonumber\\
    &\qquad +\frac{\left(\alpha-\frac12\right) }{2} \left( \frac{|\nabla T({\br})|^2}{2T({\br})} - \nabla^2 T({\br})\right)\bigg\}\bigg]\;.
    \nonumber
\end{align}
The noise probability can be re-written by inserting~\eqref{eq:od-pbp-langevin-noise-rearranged} into~\eqref{eq:od-pbp-noise-prob}, which gives in the continuous-time limit
\begin{align}
    \mathbb{P}[ \bfeta(t)] &\propto \exp \left[-\int_0^{\tf} dt \frac{\big|\dot{\br} - \big(\alpha-\frac12\big) \nabla T(\br)\big|^2}{4 T(\br)}\right]\;.
    \label{eq:od-pbp-noise-prob-rearranged}
\end{align}
Inserting~\eqref{eq:od-pbp-jacobian-result} and~\eqref{eq:od-pbp-noise-prob-rearranged} into~\eqref{eq:od-pbp-traj-prob-conversion} gives
\begin{align}
    \mathbb{P}&[\br(t)] \propto \bigg(\prod_{k=0}^{M-1} \frac{1}{(2 T(\overline{\br}^k))^{d/2}}\bigg)\label{eq:od-pbp-traj-prob-appendix}\\
    &\quad \times\exp\bigg[ \int_0^{\tf} dt \bigg\{-\frac{\big|\dot{\br} - \big(\alpha-\frac12\big) \nabla T(\br)\big|^2}{4 T(\br)}\nonumber\\
    &\quad +\frac{\left(\alpha-\frac12\right) }{2} \left( \frac{|\nabla T({\br})|^2}{2T({\br})} - \nabla^2 T({\br})\right) - \frac{\dot{\br} \cdot \nabla T({\br})}{4T({\br})}\bigg\}\bigg]\;.\nonumber
\end{align}
This result is used in the main text to prove the time-reversal symmetry of $\alpha$-discretized {OBP}s.

Note that in the main text, we use the abbreviation
\begin{align}
    \prod_{k=0}^{M-1} \frac{1}{(2 T(\overline{\br}^k))^{d/2}}\equiv F(\{\overline{\br}^k\})
\end{align}
for the time-reversal symmetric function $F$ whose exact form is unimportant. Such prefactors will be common in other calculations (e.g. Sec.~\ref{sec:trs-abp-rtp-direct-calc}) and do not impact entropy production or the detailed balance condition because they are equal to their time-reversals.

\subsection{EPR for interacting {OBP}s}\label{appendix:path-integ-pbp-int}

As in the non-interacting case, we re-write the $\alpha$-discretized Langevin equation as a Stratonovich-discretized dynamics, with an additional force $\big(\alpha-1/2\big) \nabla T(\overline{\br}^k)\Delta t$~\eqref{eq:od-pbp-ni-alpha-to-strato}. The noise probabilities are identical to the non-interacting case, but, to convert to trajectory probabilities in the $N$-particle phase space, the Jacobian must now include inter-particle forces. Thus Eq.~\eqref{eq:od-pbp-traj-prob-conversion} becomes
\begin{align}
    \mathbb{P}[\{\br_k\}] &= \mathbb{P}[\{\Delta \bfeta_k\}] \det_{\mu,\nu,i,j,k,\ell} \left(\pfrac{\Delta \eta_k^{i,\mu}}{r_\ell^{j,\nu}}\right)\;
\end{align}
where $i,j\in \{1,\ldots,N\}$, $\mu,\nu\in\{1,\ldots,d\}$, $k\in \{0,\ldots,M-1\}$, and $\ell \in \{1,\ldots,M\}$. The noise is determined by rearranging the Langevin equation as in~\eqref{eq:od-pbp-langevin-noise-rearranged}, and the elements of the Jacobian are found by direct algebra to be
\if{\begin{align}
    \pfrac{\eta_k^{i,\alpha}}{r_\ell^{j,\beta}} &= \frac{\delta^{i,j}}{\sqrt{2 \Delta T(\overline{\br}_k^i}} \Bigg\{\delta^{\alpha\beta} (\delta_{k+1,\ell} - \delta_{k,\ell})  - \frac{\Delta t}{2} \sum_j \partial^\beta f^\alpha(\overline{\br}_k^i - \overline{\br}_k^j)\nonumber\\
    &\quad + \frac12 \left(\alpha-\frac12\right) \left[\frac{\partial^\alpha T(\overline{\br}_k^i) \partial^\beta T(\overline{\br}_k^i)}{2T(\overline{\br}_k^i)}-\partial^\beta \partial^\alpha T(\overline{\br}_k^i)\right]\nonumber\\
    &\quad - \frac{\Delta \br_{k}^{i,\alpha} \partial^\beta T(\overline{\br}_k^i)}{4T(\overline{\br}_k^i)} + \frac{\Delta t \partial^\beta T(\overline{\br}_k^i)}{4T(\overline{\br}_k^i)} \sum_j f^\alpha(\overline{\br}_k^i - \overline{\br}_k^j)\Bigg\}
\end{align}}\fi
\begin{align}
    \pfrac{\Delta \eta_k^{i,\mu}}{r_\ell^{j,\nu}} &= \frac{\delta^{ij}}{\sqrt{2  T(\overline{\br}_k^i)}} \Bigg\{\delta^{\mu\nu} (\delta_{k+1,\ell} - \delta_{k,\ell}) - \frac{\Delta r_{k}^{i,\mu} \partial^\nu T(\overline{\br}_k^i)}{4T(\overline{\br}_k^i)} \nonumber\\
    &+ \Delta t F(\{\overline{\br}_k^i\})\Bigg\} + \frac{(1-\delta^{ij})}{\sqrt{2 T(\overline{\br}_k^i)}} \Delta t G(\{\overline{\br}^i_k\})\;.
\end{align}
We have absorbed the rest of the derivative into unspecified functions $F,G$ that depend only on the positions, and are thus even under time-reversal so that they cancel out when taking the ratio with the probability of the time-reverse trajectory. We note that, as in the non-interacting case, terms where $\mu\neq \nu$ or $i\neq j$ are subleading in order of $\Delta t$ with respect to the diagonal elements. Thus the determinant is found by taking the product of the diagonal terms corresponding to $\mu=\nu$, $i=j$, and $\ell=k+1$, which yields:
\begin{align}
    \det_{\mu,\nu,i,j,k,\ell} &\left(\pfrac{\Delta \eta_k^{i,\mu}}{r_\ell^{j,\nu}}\right) \\
    &= H(\{\overline{\br}_k^i\}) \exp\left[-\int_0^{\tf} dt \sum_{i=1}^N \frac{\dot{\br}^i \cdot T(\br^i)}{4T(\br^i)}\right]\;.\nonumber
\end{align}
Again, $H$ is a time-reversal symmetric term that will play no role in the following. Meanwhile, the noise probability can be rewritten in terms of spatial trajectories as
\begin{align}
    &\mathbb{P}[\{\bfeta^i(t)\}] \propto \exp \Bigg[ -\int_0^{\tf} \hspace{-0.5em} dt \sum_{i=1}^N \frac{1}{4 T(\br^i)}\\
    &\qquad \times \bigg|\dot{\br}^i - \Big(\alpha-\frac12\Big)^2 \nabla T(\br^i) - \sum_{j=1}^N \mathbf{f}_{\rm int}(\br^i-\br^j)\bigg|^2\Bigg]\nonumber
\end{align}
and the path probability is thus given by
\begin{align}
    \mathbb{P}&[\{\br^i(t)\}] = K(\{\overline{\br}_k^i,|\Delta{\br}_k^i|^2\})\\
    &\exp\Bigg[- \int_0^{\tf} dt \sum_i \Bigg\{\frac{\dot{\br}^i \cdot T(\br^i)}{4T(\br^i)}\nonumber\\
    &\quad -\frac{\dot{\br}^i \cdot \left[\big(\alpha-\frac12\big) \nabla T(\br^i) + \sum_j \mathbf{f}_{\rm int}(\br^i-\br^j)\right]}{2 T(\br^i)} \Bigg\}\Bigg] \nonumber
\end{align}
where, again, we have absorbed the time-reversal symmetric factors into an unspecified function $K$.

\section{Exact solution for non-interacting 1d RTPs}\label{appendix:rtp-1d-van-kampen}

Consider non-interacting RTPs in 1 dimension in an activity landscape $v(x)$ and force field $f(x)$, which may or may not be the gradient of a potential. Define the left- and right-moving densities $R(x,t)$ and $L(x,t)$, velocities $v_R = v(x)+f(x)$ and $v_L=v(x)-f(x)$. The equations of motion can then be written as
\begin{align}
    \dot{R} &= -\partial_x \left[v_R R\right] - \frac{R}{2{\tau}} + \frac{L}{2{\tau}}\\
    \dot{L} &= \partial_x \left[v_L L\right] + \frac{R}{2{\tau}} - \frac{L}{2{\tau}}\;.
\end{align}
The probability density can be written as $\rho=R+L$, the magnetization as $m=R-L$, and the current as $J=v_R R- v_L L$.

To determine the steady-state current and density profile, we use $\dot{\rho}=\dot{J}=0$ and re-arrange the Master equation in the spirit of what was done in~\cite{tailleur_statistical_2008,angelani_active_2011}. In the following calculation, we use  $J'=-\dot{\rho}=0$ and the identity $R-L=(J-f\rho)/v$. The steady-state condition for the current reads
\begin{widetext}
\begin{align}
    0 &= \dot{J} = v_R \dot{R} - v_L \dot{L} = -\left[v_R (v_R R)' + v_L (v_L L)'\right] - \frac{1}{2{\tau}} \left[v_R (R-L) + v_L (R-L)\right]\\
    &= -\big\{v_R^2 R' + v_L^2 L' + v_R v_R' R+v_L v_L' L\big\}  - \frac{1}{2{\tau}} (v_R+v_L)(R-L)\\
    &= -\Big\{(v^2 + f^2) \rho' + (vv'+ff')\rho  + (vf'+fv') (R-L) + 2vf (R-L)'\Big\} - \frac{1}{2{\tau}} 2v \left(\frac{J- f\rho}{v}\right)\\
    &= -\Big\{\big[(v^2+f^2)\rho\big]' - \frac12 (v^2+f^2)'\rho + f'(J-f\rho) - \frac{v' f}{v} (J-f\rho) - 2 f (f\rho)'\Big\} - \frac{(J-f\rho)}{{\tau}}\\
    &= -\Big\{\big[(v^2-f^2)\rho\big]' - \frac12 (v^2)'\rho + \frac{f^2 v'}{v} \rho + J\Big( f' - \frac{v' f}{v}\Big)\Big\} - \frac{(J-f\rho)}{{\tau}}
\end{align}
\end{widetext}
which we can solve for $J$ to find
\begin{align}
    J &= \frac{\Big\{\left[{\tau}(f^1-v^2)\rho\right]' + \rho \big( f + \frac{(v^2)' {\tau}}{2 } - f^2 \frac{v' {\tau}}{v } \big)\Big\}}{1-\frac{v' f {\tau}}{v }+{\tau} f'}\;.\label{eq:rtp-1d-U-v-J-ss}
\end{align}
One can immediately see the similarities between Eqs.~\eqref{eq:rtp-1d-U-v-J-ss} and~\eqref{eq:fpe-1d-pbp-van-kampen}, which describes the current for an {OBP} in an inhomogeneous medium (Sec.~\ref{sec:van-kampen-review-1d-fokker-planck}). We can thus read off the effective fields
\begin{align}
    \mu_\eff &= \frac{1}{1-\frac{v' f {\tau}}{v}+{\tau} f'}\label{eq:rtp-1d-U-v-mueff} \; ,\\
    F_\eff &= f + \frac{(v^2)' {\tau}}{2 } - f^2 \frac{v' {\tau}}{v } \label{eq:rtp-1d-U-v-Feff} \; ,\\
    T_\eff &= {\tau}(v^2-f^2)\;.\label{eq:rtp-1d-U-v-Teff}
\end{align}
Solving for the steady-state density and current for this system amounts to plugging $\mu_\eff,\,F_\eff,$ and $T_\eff$ into Eqs.~\eqref{eq:VKsol},~\eqref{eq:VKsolJ}, and~\eqref{eq:od-pbp-van-kampen-phi-def}. The pseudo potential is then given by:
\begin{align}
    \Phi(x)&= \int_0^x du \frac{\frac{f(u)^2 v'(u)}{v(u)} - \frac{f(u)}{{\tau}} - v'(u) v(u)}{v(u)^2-f(u)^2}\;,\label{eq:rtp-1d-v-U-Phi-sol}
\end{align}
while the density and current read
\begin{widetext}
    \begin{align}
        \rho(x) &= \frac{e^{-\Phi(x)}}{v(x)^2-f(x)^2} \frac{1}{{\tau}} \left[\rho(0) {\tau} \left(v_0^2 - f(0)^2\right) - J \int_0^x e^{\Phi(u)} \left(1-\frac{v'(u) f(u) {\tau}}{v(u)}+f'(u){\tau}\right) du\right] \; ,\label{eq:rtp-1d-v-U-rho-sol}\\
        J &= \frac{1-e^{\Phi(L)}}{\int_0^L \frac{dx }{{\tau}(v(x)^2-f(x)^2)} \int_0^L dx' \left(1-\frac{v'(x') f(x') {\tau}}{v(x')}+{\tau} f'(x)\right) e^{\Phi(x')-\Phi(x)} \left[1 + \Theta(x-x') \left(e^{\Phi(L)}-1\right)\right]}\label{eq:rtp-1d-v-U-J-sol}.
    \end{align}
\end{widetext}

\begin{figure}
    \centering
    \begin{tikzpicture}
    \path (-0,0) node {\includegraphics[width=8.7cm]{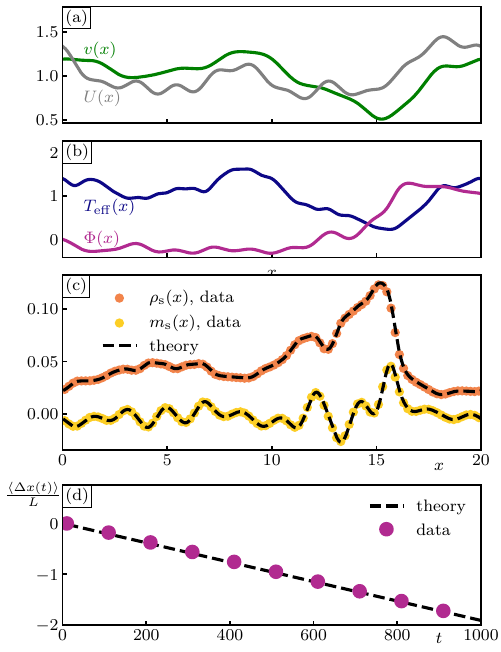}};
    \end{tikzpicture}
    \caption{\textbf{Particle simulations of non-interacting RTPs in $d=1$.} {\bf (a)} The activity and potential landscapes are given by $v(x)$ and $U(x)$, respectively. {\bf (b)} The effective temperature $T_{\rm eff}(x)$ and pseudo-potential $\Phi(x)$, as defined in Eqs.~\eqref{eq:rtp-1d-U-v-Teff} and~\eqref{eq:rtp-1d-v-U-Phi-sol}. {Note the aperiodicity of $\Phi(x)$.} {\bf (c)} Particle density $\rho$ and polarization $m$, along with the exact solution (black), whose explicit form for $\rho$ is written in Eq.~\eqref{eq:rtp-1d-v-U-rho-sol}. {\bf (d)} Average particle displacement over time (points), along with the theoretical prediction (line) given in Eq.~\eqref{eq:rtp-1d-v-U-J-sol}. Simulation details are given in Appendix~\ref{appendix:sim-abp-rtp}.}
    \label{fig:RTP-U-v-exact}
\end{figure}

If $f(x)=0$, we see that $\Phi(x) = -\ln \big[\frac{v(x)}{v(0)}\big]$ and therefore $\Phi(L)=0$ and $J=0$. 

If $f(x)\neq 0$ but $v(x)$ is a constant $v_0$ (a case that has received considerable attention, e.g. in~\cite{angelani_active_2011,zhen_optimal_2022,roberts_run-and-tumble_2023}), the condition to have a nonzero current reads
\begin{align}
    0 \neq -\Phi(L) = \frac{1}{v_0^2 {\tau}} \int_0^L dx \frac{\Tilde{f}(x)}{1 - \Tilde{f}(x)^2} \; ,
\end{align}
where $\Tilde{f}(x)=f(x)/v_0$. The sign of $J$ is the same as the sign of $-\Phi(L)$, which is helpful to determine the sign of the current for RTPs in complex potential landscapes, e.g. in quenched random potentials.
Note that in regions where $\Tilde{f}(x)<1$ the particles are unable to move in the direction opposite to $f$, and can be trapped in a potential minimum.
The solution doesn't apply to systems where such losses of ergodicity occur.

In the general case where $f(x)$ and $v(x)$ vary spatially, as originally studied for ABPs in~\cite{pototsky_rectification_2013}, the condition for a finite {directed} current---namely,  the aperiodicity of $\Phi(x)$ in Eq.~\eqref{eq:rtp-1d-v-U-Phi-sol}---specifies the joint condition for $V(x)$ and $v(x)$ to induce a current. 

This exact solution agrees with numerical particle-based simulations, as demonstrated in Fig.~\ref{fig:RTP-U-v-exact}.

\section{EPR of interacting AOUPs in activity landscapes}
\label{appendix:path-integral-aoup}
Here we detail the calculation of the EPR for interacting AOUPs in an activity landscape given in Eq.~\eqref{eq:aoup-epr-density}. This calculation is quite similar to that of {UBP}s given in Sec.~\ref{appendix:path-integ-upbp-int}; however, there are important differences, and we will thus detail it separately here. As for {UBP}s, although the AOUP dynamics~\eqref{eq:langevin-aoup-a}-\eqref{eq:langevin-aoup-b} suffer no discretization ambiguity, it is necessary to use caution when calculating path probabilities. Thus we consider a Stratonovich-discretized version of the dynamics, using $\overline{\br}_i^k$, the position evaluated at the midpoint of the $k$th timestep, as defined in Eq.~\eqref{eq:overline-modpoint-strato-definition}, and the spatial increments $\Delta \br_i^k$ as defined in Eq.~\eqref{eq:spatial-increment-deltar}. We thus find the discretized dynamics:
\begin{align}
    \tau& \left(\frac{\Delta \br_i^{k+1}-\br_i^k}{\Delta t^2}\right) = - \frac{\Delta \br_i^k}{\Delta t} + \sqrt{2 D(\overline{\br}_i^k)} \Delta \bfeta_i^{k+1}  \label{eq:aoup-discretized-langevin}\\
    &\quad +\sum_j \left[1 + \tau \left(\frac{\Delta \br_i^k}{\Delta t} - \frac{\Delta \br_j^k}{\Delta t}\right)\cdot \nabla \right] \bfint(\overline{\br}_i^k-\overline{\br}_j^k)\nonumber
\end{align}
where $\Delta \bfeta_i^{k+1}$ is a centered Gaussian noise that satisfies $\langle \Delta \eta_{i,\mu}^k \Delta \eta_{j,\nu}^\ell\rangle = \delta_{ij} \delta_{\mu\nu} \delta_{k\ell}/\Delta t$. We know the probability of a realization of this noise to be given by
\begin{align}
    \mathbb{P}[\{\Delta \bfeta_i^k\}|\{\br_i^0,\br_i^1\}] &\propto \exp\left[-\sum_{i=1}^N \sum_{k=1}^{M-1} \frac{\Delta t}{2} |\Delta \bfeta_i^k|^2\right]\;,\label{eq:aoup-noise-prob}
\end{align}
which can be used to find the probability of a spatial trajectory, using the relation
\begin{align}\label{eq:probatrajaoup}
    \mathbb{P}[\{\br_i^k\}|\{\br_i^0,\br_i^1\}]&= \mathbb{P}[\{\Delta \bfeta_i^k\}|\{\br_i^0,\br_i^1\}] \left|\frac{\mathbb{D}[\{\Delta \bfeta_i^k\}|\{\br_i^0,\br_i^1\}]}{\mathbb{D}[\{\br_i^k\}|\{\br_i^0,\br_i^1\}]}\right|\;.
\end{align}
Here, $\big|{\mathbb{D}[\{\Delta \bfeta_i^k\}\big|\{\br_i^0,\br_i^1\}]}/{\mathbb{D}[\{\br_i^k\}\big|\{\br_i^0,\br_i^1\}]}\big|$ is the Jacobian between the $M-1$ variables $\{\Delta \bfeta_i^k\}_{k=1}^{M-1}$ and the $M-1$ variables $\{\br_i^k\}_{k=2}^M$. Note the shift in $k$ indices: $\Delta \bfeta_i^k$ determines $\br_i^{k+1}$. Note that we condition on the positions in the first two timesteps, $\{\br_i^0,\br_i^1\}$, which is equivalent to conditioning on the initial positions and momenta. We first note, using the relation between the variables given by Eq.~\eqref{eq:aoup-discretized-langevin}, that this Jacobian will be upper triangular in the time dimension, because 
\begin{align}
    \pfrac{\Delta \eta_{i,\mu}^{k+1}}{r_{j,\nu}^\ell} = 0 \;\;\text{whenever}\;\;\ell>k+2\;.
\end{align}
For the diagonal elements $\partial \Delta \eta_{i,\mu}^{k+1}/\partial r_{j,\nu}^{k+2}$, we have
\begin{equation}
    \frac{\delta_{ij} \delta_{\mu\nu} \tau}{\Delta t^2} = \sqrt{2 D(\overline{\br}_i^k)} \pfrac{\Delta \eta_{i,\mu}^{k+1}}{r_{j,\nu}^{k+2}}
\end{equation}
from which we find the determinant
\begin{align}
    \left|\frac{\mathbb{D}[\{\Delta \bfeta_i^k\}\big|\{\br_i^0,\br_i^1\}]}{\mathbb{D}[\{\br_i^k\}\big|\{\br_i^0,\br_i^1\}]}\right| &= \prod_{k=0}^{M-2} \prod_{i=1}^N \frac{\tau}{\Delta t^2 \sqrt{2 D(\overline{\br}_i^k)}} \label{eq:aoup-jac-det}\\
    &\propto \prod_{k=0}^{M-2} \prod_{i=1}^N \frac{1}{\sqrt{D(\overline{\br}_i^k)}}\;.
\end{align}
Next, we convert the noise probability [Eq.~\eqref{eq:aoup-noise-prob}] into spatial coordinates using Eq.~\eqref{eq:aoup-discretized-langevin}:
\begin{align}\label{eq:aoup-trajfinal}
    \mathbb{P}&[\{\Delta \bfeta_i^k\}\big|\{\br_i^0,\br_i^1\}] \\
    &\propto \exp\bigg[-\sum_{i=1}^N \sum_{k=0}^{M-2} \frac{\Delta t }{4 D(\overline{\br}_i^k)} \bigg|\tau \left(\frac{\Delta \br_i^{k+1}-\Delta \br_i^k}{\Delta t^2}\right) \nonumber\\
    &\quad + \frac{\Delta \br_i^k}{\Delta t} - \sum_{j=1}^N \Big[1 + \tau \Big(\frac{\Delta \br_i^k}{\Delta t}-\frac{\Delta \br_j^k}{\Delta t}\Big)\cdot \nabla\Big] \bfint(\overline{\br}_i^k - \overline{\br}_j^k)\bigg|^2\bigg]\nonumber\;.
\end{align}
Equations~\eqref{eq:probatrajaoup},~\eqref{eq:aoup-jac-det}, and~\eqref{eq:aoup-trajfinal} give the trajectory probability density. Once again defining a trajectory's time-reverse as $\br_i^{R,k} \equiv \br_i^{M-k}$, we find the ratio between the forward and reverse trajectory probabilities to be
\begin{widetext}
\begin{align}
    &\frac{\mathbb{P}[\{\br_i^k\}\big|\{\br_i^0,\br_i^1\}]}{\mathbb{P}[\{\br_i^{R,k}\}\big|\{\br_i^{R,0},\br_i^{R,1}\}]} \notag\\
    &= \sqrt{\frac{D(\overline{\br}_i^{M-1})D(\overline{\br}_i^{M-2})}{D(\overline{\br}_i^0)D(\overline{\br}_i^1)}}
     \exp\Bigg[-\sum_{i=1}^N \sum_{k=0}^{M-2} \frac{\Delta t}{4} \bigg\{\frac{1}{ D(\overline{\br}_i^k)}  \bigg|\tau \frac{\Delta \br_i^{k+1}-\Delta \br_i^k}{\Delta t^2} + \frac{\Delta \br_i^k}{\Delta t} - \sum_{j=1}^N  \Big[1 + \tau \Big(\frac{\Delta \br_i^k}{\Delta t}-\frac{\Delta \br_j^k}{\Delta t}\Big)\cdot \nabla\Big] \bfint(\overline{\br}_i^k - \overline{\br}_j^k)\bigg|^2\nonumber\\
    &\quad - \frac{1}{D(\overline{\br}_i^{M-k-1})} \bigg|\tau\frac{\Delta \br_i^{M-k-1}-\Delta \br_i^{M-k-2}}{\Delta t^2} 
    - \frac{\Delta \br_i^{M-k-1}}{\Delta t} - \sum_{j=1}^N \Big[1 - \Big(\frac{\Delta \br_i^{M-k-1}}{\Delta t}-\frac{\Delta \br_j^{M-k-1}}{\Delta t}\Big)\cdot \nabla\Big] \bfint(\overline{\br}_i^{M-k-1} - \overline{\br}_j^{M-k-1})\bigg|^2\bigg\}\Bigg]\nonumber\\
    &= \sqrt{\frac{D(\overline{\br}_i^{M-1})D(\overline{\br}_i^{M-2})}{D(\overline{\br}_i^0)D(\overline{\br}_i^1)}} \exp\bigg[-\sum_{i=1}^N \sum_{k=0}^{M-2} \frac{\Delta t }{4 D(\overline{\br}_i^k)}  \bigg|\tau\frac{\Delta \br_i^{k+1}-\Delta \br_i^k}{\Delta t^2} + \frac{\Delta \br_i^k}{\Delta t} - \sum_{j=1}^N \Big[1 + \tau \Big(\frac{\Delta \br_i^k}{\Delta t}-\frac{\Delta \br_j^k}{\Delta t}\Big)\cdot \nabla\Big] \bfint(\overline{\br}_i^k - \overline{\br}_j^k)\bigg|^2\nonumber\\
    &\quad + \sum_{i=1}^N \sum_{k=1}^{M-1}  \frac{\Delta t }{4 D(\overline{\br}_i^k)} \bigg|\tau \frac{\Delta \br_i^k-\Delta \br_i^{k-1}}{\Delta t^2} - \frac{\Delta \br_i^k}{\Delta t} -\sum_{j=1}^N \Big[1 - \tau \Big(\frac{\Delta \br_i^k}{\Delta t}-\frac{\Delta \br_j^k}{\Delta t}\Big)\cdot \nabla\Big] \bfint(\overline{\br}_i^k - \overline{\br}_j^k)\bigg|^2\bigg]\;,\label{eq:noname-aoup}
\end{align}
\end{widetext}
where, in the second equality, we have relabelled the time indices for the backward path as $M-k-1 \to k$. In the last line of Eq.~\eqref{eq:noname-aoup}, we may replace $\frac{\Delta \br_i^k-\Delta \br_i^{k-1}}{\Delta t^2}$ with $\frac{\Delta \br_i^{k+1}-\Delta \br_i^k}{\Delta t^2}$, since their difference is of order $\mathcal{O}(\sqrt{\Delta t})$ and thus do not contribute to the sum in the limit $\Delta t \to 0$. Taking the natural logarithm of Eq.~\eqref{eq:noname-aoup} then leads to the following result for the entropy production $\Sigma$ along the path:
\begin{widetext}
\begin{align}
    \Sigma(\tf) \equiv& \ln \bigg[\frac{\mathbb{P}[\{\br_i^k\}\big|\{\br_i^0,\br_i^1\}]}{\mathbb{P}[\{\br_i^{R,k}\}\big|\{\br_i^{R,0},\br_i^{R,1}\}]}\bigg]\\
    =& -\bigg\{ \sum_{i=1}^N \sum_{k=1}^{M-2} \frac{\Delta t}{D(\overline{\br}_i^k)}  \bigg[\tau \frac{\Delta \br_i^{k+1}-\Delta \br_i^k}{\Delta t^2}- \sum_{j=1}^N \bfint(\overline{\br}_i^k - \overline{\br}_j^k)\bigg] \cdot \bigg[\frac{\Delta \br_i^k}{\Delta t} - \tau \sum_{j=1}^N \left(\frac{\Delta \br_i^k}{\Delta t}-\frac{\Delta \br_j^k}{\Delta t}\right) \cdot \nabla \bfint(\overline{\br}_i^k-\overline{\br}_j^k)\bigg] \bigg\}\nonumber \\
    & +\frac12 \ln \bigg[\frac{D(\overline{\br}_i^{M-1})D(\overline{\br}_i^{M-2})}{D(\overline{\br}_i^0)D(\overline{\br}_i^1)}\bigg]\label{eq:aoup-boundaryterms}+\mathcal{O}(\sqrt{\Delta t})\nonumber\\
    \underset{\Delta t\rightarrow 0}{\longrightarrow}& - \sum_{i=1}^N \int_0^\tf dt \frac{1 }{D(\br_i)} \bigg[\tau \ddot{\br}_i - \sum_{j=1}^N \bfint(\br_i-\br_j)\bigg] \cdot \bigg[\dot{\br}_i - \tau \sum_{j=1}^N (\dot{\br}_i-\dot{\br}_j) \cdot \nabla \bfint(\br_i-\br_j)\bigg] + \ln \left[\frac{D(\overline{\br}_i(\tf))}{D(\overline{\br}_i(0))}\right] \;.
\end{align}   
\end{widetext}
In Eq.~\eqref{eq:aoup-boundaryterms}, the $\mathcal{O}(\sqrt{\Delta t})$ term corresponds to the missing $k=0$ and unmatched $k=M-1$ terms that appear in the last line of Eq.~\eqref{eq:noname-aoup}. They do not contribute in the limit $\Delta t\to 0$.
The EPR $\sigma$ is then defined as $    \sigma = \lim_{\tf\rightarrow\infty} \frac{\Sigma(\tf)}{\tf}$ and we thus find:
\begin{align}
    \sigma &= \sum_{i=1}^N \Bigg\langle\frac{ 1}{D(\br_i)} \bigg[\sum_{j=1}^N \bfint(\br_i-\br_j) - \tau\ddot{\br}_i\bigg]\notag \\
    &\quad\cdot \bigg[\dot{\br}_i - \tau\sum_{j=1}^N (\dot{\br}_i-\dot{\br}_j)\cdot \nabla \bfint(\br_i-\br_j)\bigg]\Bigg\rangle\;,\label{eq:aoups-sigmahat}
\end{align}
where we have assumed the system to be ergodic. In Eq.~\eqref{eq:aoup-epr-density}, we then define a fluctuating EPR density $\hat{\sigma}(\br)$ such that $\int d\br \langle \hat{\sigma}(\br)\rangle=\sigma$. The term proportional to $\tau$ in Eq.~\eqref{eq:aoups-sigmahat} is specific to AOUPs and distinguishes their EPR from that of {UBP}s.

{
\section{Additional examples of blowtorch ratchets}\label{app:blowtorch}

\begin{figure*}
    \centering
    \includegraphics[width=7in]{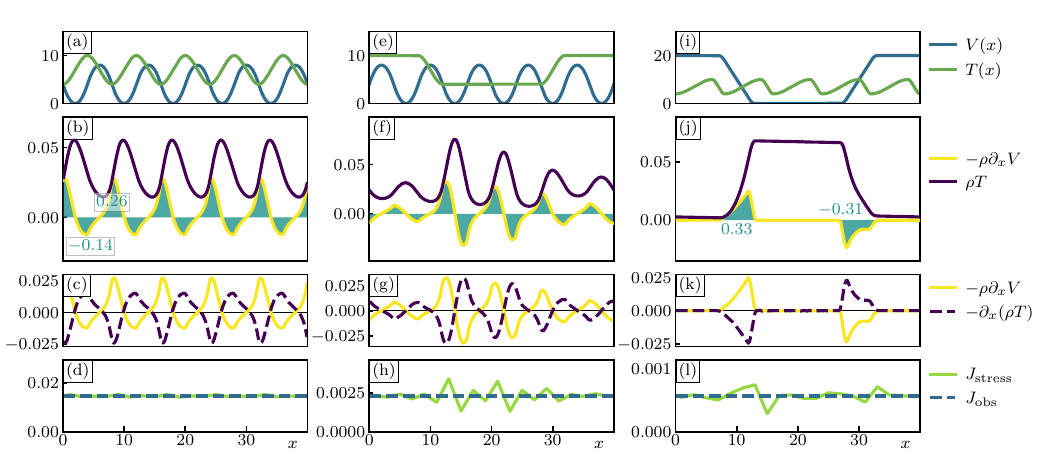}
    \caption{{\textbf{Momentum sources in the ``blowtorch ratchet."} Simulations of interacting {OBP}s in an external potential with spatially-varying potential and temperature. 
    {\bf (a), (e), (i)} Temperature (green) and external potential (blue) used in the simulations. {\bf (b), (f), (j)} Thermal stress $\rho(x) T(x)$ (dark purple) and momentum source $-\rho(x) \partial_x V(x)$ (yellow) measured in simulations. The momentum exchanged with the potential is shaded in blue. The momentum sources do not balance only when $T(x)$ is spatially-varying.
    {\bf (c), (g), (k)} The contributions to the current from the stress (dashed purple) and from the momentum source (yellow). {\bf (d), (h), (l)} $J_{\rm stress} = -\rho \partial_x V - \partial_x (\rho T)$, the current due to stress and momentum sources (dashed teal), agrees with the current $J_{\rm obs}$ measured from particle displacements. Simulation details are given in Appendix~\ref{appendix:sim-blowtorch}.}}
    \label{fig:blowtorch-app}
\end{figure*}

In this Appendix, we demonstrate how other blowtorch ratchets can be understood through the rectification of the momentum sources provided by an external potential via temperature variations, which do not themselves directly inject any momentum. It is known~\cite{van_kampen_relative_1988} that in a system of {OBP}s with external potential $V(x)$ and spatially-varying temperature $T(x)$, the ratchet current is nonzero if and only if
\begin{align}
    \Phi(L) &= \int_0^L dx \frac{\partial_x V(x)}{T(x)} \neq 0\;.
\end{align}
Evidently, there is an endless variety of possible $V(x)$ and $T(x)$ that satisfy this requirement. In Fig.~\ref{fig:blowtorch-app}, we demonstrate three such possibilities, all of which display currents.

In panels~(a-d), we show the results of a temperature and force landscape which are both parity-symmetric but phase shifted with respect to each other. This generates a sequence of alternating sources of left and right momentum, shown in panel (b), which do not balance. In panel (d), we show that the sum of these momentum sources and the stress gradients ($J_{\rm stress}$) directly accounts for the current measured by directly tracking particles ($J_{\rm obs}$). 

In panels~(e-h) of  Fig.~\ref{fig:blowtorch-app}, we show another system where, similarly, a parity-symmetric $V(x)$ and $T(x)$ can produce a current thanks to their relative asymmetry. This time, we use a step-like $T(x)$ which cubically interpolates between two values. The alternating momentum sources are now globally modulated by the value of $T(x)$, but due to differences in the force landscape between the two regions of varying temperature, this modulation is asymmetric, and the sum of all momentum sources is nonzero, leading to a current.

Finally, in panels~(i-l) of Fig.~\ref{fig:blowtorch-app}, we show how an asymmetric temperature landscape $T(x)$, though unable to do so on its own, can lead to steady currents in the presence of an external potential.
}

\bibliography{refs-resub}

\end{document}